\documentclass[%
pre,
superscriptaddress,
tightenlines,
showpacs,showkeys,
a4paper,12pt
]{revtex4}
\usepackage{dcolumn}
\usepackage{amssymb,amsmath,stmaryrd,array}

\usepackage{graphicx}
\usepackage{subfig}

\makeatletter

\newcommand{\avr}[1]{\ensuremath{\langle{#1}\rangle}}

\newcommand{\cnj}[1]{{#1}^{\ast}}

\newcommand{\tcnj}[1]{{#1}^{T}}

\newcommand{\pdrs}[1]{\partial_{#1}}
\newcommand{\pdr}[2]{\dfrac{\partial #1}{\partial #2}}
\newcommand{\drf}[2]{\dfrac{\dd #1}{\dd #2}}
\newcommand{\vdr}[2]{\dfrac{\delta #1}{\delta #2}}

\newcommand{\bnbl}{\boldsymbol{\nabla}}

\newcommand{\diag}{\mathop{\rm diag}\nolimits}

\newcommand{\Tr}{\mathop{\rm Tr}\nolimits}

 \newcommand{\bs}[1]{\boldsymbol{#1}}
 \newcommand{\vc}[1]{\mathbf{#1}}
 \newcommand{\mvc}[1]{\mathbf{#1}}
 \newcommand{\uvc}[1]{\hat{\mathbf{#1}}}

\newcommand{\dd}{\mathrm{d}}

\newcommand{\JJ}{\ensuremath{\mathcal{J}}}
 
 \newcommand{\ee}{\mathrm{e}}

\newcommand{\st}{\mathrm{st}}

\newcommand{\av}{\mathrm{av}}

\newcommand{\trans}{\textit{trans}}
\newcommand{\cis}{\textit{cis}}
\newcommand{\grnd}{\mathrm{G}}
\newcommand{\exct}{\mathrm{E}}

\newcommand{\inc}{\mathrm{inc}}
\newcommand{\refl}{\mathrm{refl}}
\newcommand{\transm}{\mathrm{tr}}
\newcommand{\vac}{\mathrm{vac}}
\newcommand{\med}{\mathrm{m}}

\makeatother

\begin{document}
\DeclareGraphicsExtensions{.eps,.png,.pdf}
\title{
 Kinetics of photoinduced ordering in azo-dye films:\\
two-state and diffusion models
}

\author{Alexei~D.~Kiselev}
\email[Email address: ]{kiselev@iop.kiev.ua}

\affiliation{%
Hong Kong University of Science and Technology,
Clear Water Bay, Kowloon,
Hong Kong}

\affiliation{%
 Institute of Physics of National Academy of Sciences of Ukraine,
 prospekt Nauki 46,
 03028 Ky\"{\i}v, Ukraine}

\author{Vladimir~G.~Chigrinov}
\email[Email address: ]{eechigr@ust.hk}

\author{Hoi-Sing~Kwok}
\email[Email address: ]{eekwok@ust.hk}

\affiliation{%
Hong Kong University of Science and Technology,
Clear Water Bay, Kowloon,
Hong Kong}

\date{\today}

\begin{abstract}
We theoretically study the kinetics of photoinduced ordering in azo-dye
photoaligning layers and present the results of modeling performed
using two different phenomenological approaches.  A phenomenological
two state model is deduced from the master equation for the
one-particle distribution functions of an ensemble of two-level
molecular systems by specifying the angular redistribution
probabilities and by expressing the order parameter correlation
functions in terms of the order parameter tensor.  Using an
alternative approach that describes light induced reorientation of
azo-dye molecules in terms of a rotational Brownian motion, we
formulate the two-dimensional (2D) diffusion model as the free energy
Fokker-Planck equation simplified for the limiting regime of purely
in-plane reorientation.
The models are employed to interpret the irradiation time dependence
of the absorption order parameters defined in terms of the the
principal extinction (absorption) coefficients.
Using the exact solution to the light transmission problem for a
biaxially anisotropic absorbing layer, these coefficients are
extracted from the absorbance-vs-incidence angle curves measured at
different irradiation doses for the probe light linearly polarized
parallel and perpendicular to the plane of incidence.
It is found that, in the azo-dye films, the transient photoinduced
structures are biaxially anisotropic whereas the photosteady and the
initial states are uniaxial.
\end{abstract}

 \maketitle

\section{Introduction}
\label{sec:intro}

It has long been known that
some photosensitive materials
such as compounds containing azobenzene and its derivatives
may become dichroic and
birefringent under the action of light.
This phenomenon~---~the so-called effect
of  \textit{photoinduced optical 
anisotropy} (POA)~---~has a long history dating 
back almost nine decades
to the paper by Weigert~\cite{Wieg:1919}.

The Weigert effect (POA)
has been attracted much attention over
the past few decades because
of its technological importance  
in providing tools to produce the light-controlled anisotropy.
For example, the materials that
exhibit POA are very promising for use in many
photonic applications~\cite{Eich:1987,Nat:1992,Pras:1995,Blinov:chemph:1999}.

It is also well known that producing substrates with anisotropic anchoring
properties is one of the key procedures in the fabrication of liquid crystal
electrooptic devices.
The traditional method
widely used to align liquid crystal display cells involves mechanical
rubbing of aligning layers and has a number of the well known 
difficulties~\cite{Chigr:1999}.
The \textit{photoalignment} technique 
suggested in Refs.~\cite{Gibbon:nat:1991,Chig:jjap:1992,Dyad:jetpl:1992}
is an alternative method 
that avoids the drawbacks of the mechanical surface treatment
by using linearly
polarized ultraviolet (UV) light to induce anisotropy of the angular
distribution of molecules in a photosensitive film~\cite{Kelly:jpd:2000,Chigrin:bk:2008}.
Thus the phenomenon of POA (the Weigert effect) is at the heart of 
the photoalignment method. 


Light induced ordering  in photosensitive materials, 
though not being understood very well,
can generally occur by a variety of photochemically induced
processes.  
These typically may involve such transformations as photoisomerization, crosslinking,
photodimerization and photodecomposition 
(a recent review can be found in Ref.~\cite{Chigr:rewiev:2003,Chigrin:bk:2008}).

So, the mechanism 
underlying POA and its properties cannot be universal.
Rather they crucially depend on the material in question
and on a number of additional factors such as
irradiation conditions, surface interactions etc. 
In particular, these factors combined with the action of light
may result in 
different regimes of the photoinduced ordering kinetics
leading to the formation of various photoinduced orientational
structures (uniaxial, biaxial, splayed).  

POA was initially studied 
in viscous solutions of azodyes~\cite{Nep:1963} 
and in azodye-polymer blends~\cite{Tod:1984},
where the anisotropy was found to be  rather unstable.
This is the case where 
the photoinduced anisotropy disappears after switching off 
the irradiation~\cite{Nep:1963,Tod:1984,Dum:1993,Dum:1996,Sekkat:jpcb:2002,Raschella:josa:2007}. 
By contrast to this case, POA can be long term stable.

The stable POA was
observed in polymers containing chemically linked
azochromophores (azopolymers)~\cite{Eich:1987}.  
It turned out that
stable anisotropy can be induced in both amorphous and liquid
crystalline (LC)
azopolymers~\cite{Eich:1987,Nat:1992,Holme:1996,Petry:1993,Wies:1992,Blin:1998,Kis:epj:2001}.

The photoalignment has also been studied in a number of
similar polymer systems including dye doped polymer layers~\cite{Gibbon:nat:1991,Furum:1999},
cinnamate polymer
derivatives~\cite{Chig:jjap:1992,Dyad:jetpl:1992,Gal:1996,Barn:2000}
and side chain
azopolymers~\cite{Petry:1993,Holme:1996,Blin:1998,Iked:2000}.
In addition, the films containing photochemically stable azo dye
structures (azobenzene sulfuric dyes) were recently investigated as  new
photoaligning materials for 
nematic liquid crystal (NLC) 
cells~\cite{Chig:lc:2002,Chig:pre:2003,Kis:pre2:2005}.

In Ref.~\cite{Chig:pre:2003}, 
it was found that, owing to high degree of the photoinduced ordering, 
these films used as aligning substrates are characterized by 
the anchoring energy strengths comparable to the rubbed polyimide films.
For these materials,  the voltage holding ratio and thermal stability
of the alignment turned out to be high. 
The azo-dye films are thus promising materials for applications in
liquid crystal devices. 

According to Ref.~\cite{Kis:pre2:2005}, the anchoring characteristics 
of the azo-dye films such as the polar and azimuthal anchoring energies 
are strongly influenced by the photoinduced ordering.
In this paper the kinetics of such ordering will be of our primary
interest. 
More specifically, we deal with theoretical approaches
and related phenomenological models
describing how amount of the photoinduced
anisotropy characterized by absorption dichroism evolves 
in time upon illumination and after switching it off.

There are a number of 
models~\cite{Ped:1997,Ped:1998,Puch:1998,Hvil:2001,Kis:epj:2001,Kis:jpcm:2002,Sekkat:jpcb:2002} 
formulated  for 
azocompounds exhibiting POA driven by the \trans-\cis\ photoisomerization.
Generally, in these models, 
a sample is treated as
an ensemble of two level molecular systems:
the stable \trans\ isomers characterized by
elongated rod-like molecular conformation
can be regarded as the ground state molecules
whereas the bent banana-like shaped 
\cis\ isomers are represented by 
the excited molecules.

The photoisomerization mechanism assumes 
that the key processes behind the orientational ordering
of azo-dye molecules 
are photochemically induced \trans-\cis\  isomerization and subsequent
thermal and/or photochemical \cis-\trans\  back isomerization of
azobenzene chromophores. 

Owing to pronounced absorption dichroism of photoactive groups, 
the rate of the photoinduced isomerization strongly
depends on orientation of the azo-dye molecules relative
to the polarization vector of the actinic light, $\vc{E}_{UV}$. 
Since the optical transition dipole moment is
approximately directed along the long molecular axis,
the molecules oriented perpendicular to $\vc{E}_{UV}$ 
are almost inactive.

When the \cis\ isomers are short-living, 
the \cis\ state becomes
temporary populated during photoisomerization but reacts immediately
back to the stable \trans\ isomeric form.
The
\trans-\cis-\trans\ isomerization cycles are accompanied by 
rotations of the azo-dye molecules that tend to 
minimize the absorption and 
become oriented along directions normal to 
the polarization vector of the exciting light
$\vc{E}_{UV}$. 
Non-photoactive groups may then undergo reorientation due to
cooperative 
motion~\cite{Holme:1996,Nat:1998,Puch:1998,Kis:jpcm:2002,Sekkat:jpcb:2002}.

The above scenario, initially suggested in Ref.~\cite{Nep:1963},  
is known as the regime of \textit{photoorientation (angular redistribution)} 
where the lifetime of  \cis\ isomers is short
and POA is mainly due to the angular redistribution  of the long axes of
the \trans\ molecules during the \textit{trans--cis--trans}
photoisomerization cycles. 
Note that, in the opposite case
of long-living \cis\ isomers,
the regime of \textit{angular hole burning (photoselection)}
occurs so that
the anisotropy is caused by angular selective
burning of mesogenic \trans\  isomers 
due to stimulated transitions to
non-mesogenic \cis\  form~\cite{Dum:spie:1994,Blin:1998,Kis:jpcm:2002}. 

From the above it might be concluded that, 
whichever regime of the ordering takes place, 
the photoinduced orientational structure 
results from preferential alignment
of azo-dye molecules along the directions perpendicular to the
polarization vector of the actinic light, $\vc{E}_{UV}$,
determined by the dependence of the photoisomerization rate
on the angle between $\vc{E}_{UV}$ and the long molecular axis.
So, it can be expected that the structure will be uniaxially anisotropic with 
the optical axis directed along the polarization vector.

Experimentally, this is, however, not the case.  
For example, constraints imposed by a medium may suppress
out-of-plane reorientation of the azobezene chromophores
giving rise to the structures with 
strongly preferred in-plane alignment~\cite{Kis:epj:2001}.
Another symmetry breaking effect induced by polymeric environment 
is that the photoinduced orientational structures can be
biaxial~\cite{Wies:1992,Buff:1998,Kis:cond:2001,Kis:epj:2001,Kis:jpcm:2002,Kis:pre:2003}
(a recent review concerning medium effects on photochemical processes
can be found in~\cite{Ramam:inbk:2005}).

It was recently found that,
similar to the polymer systems, 
the long-term stable POA in the azo-dye SD1 films
is characterized by the biaxial photoinduced structures with favored 
in-plane alignment~\cite{Kiselev:idw:2008}.
Unlike azopolymers, photochromism in  
these films is extremely weak 
so that it is very difficult to unambiguously detect the presence of
a noticeable fraction of \cis\ isomers. 

As compared to the polymer systems, 
modeling of photoinduced ordering in the azo-dye films has received little attention.
In this paper we intend to fill the gap
and describe the symmetry breaking and biaxiality effects
using phenomenological models formulated 
on the basis of a unified approach to the kinetics of 
POA~\cite{Kis:epj:2001,Kis:jpcm:2002}.
The layout of the paper is as follows.

In Sec.~\ref{subsec:order-param},
we introduce necessary notations and 
discuss the relationship between
the order parameter and the absorption tensors.
Then, in Sec.~\ref{subsec:master-equation},
we recapitulate the theory~\cite{Kis:jpcm:2002} 
by assuming that the azo-dye molecules can be
represented by two level molecular systems.
This theoretical approach is based on 
the master equation combined with the kinetic equation for 
the additional (matrix) system, 
which phenomenologically accounts for 
the presence of long-living anisotropic (angular) 
correlations.

In Sec.~\ref{subsec:2-lvl-model},
a phenomenological two state model is introduced by specifying 
the angular redistribution probabilities and 
by expressing the order
parameter correlation functions in terms of the order parameter
tensor.
In this model, the regime of photoorientation with 
short living excited molecules is characterized by 
weak photochromism and
negligibly small fraction of \cis\ isomers that 
rapidly decays after switching off irradiation.

According to Ref.~\cite{Chig:pre:2004},
when the photochemical processes underlying photoisomerization 
are hindered,
the process of photoinduced reorientation can be 
alternatively described 
as rotational diffusion of azo-dye molecules 
under the action of the polarized light. 

In Sec.~\ref{subsec:mf-planck-eq},
we show that diffusion models of POA
can be formulated as the free energy Fokker-Planck 
equation~\cite{Frank:bk:2005} 
describing light induced reorientation of
azo-dye molecules as rotational Brownian motion 
governed by the effective mean field potential.
Using this approach,
the diffusion model suggested in~\cite{Chig:pre:2004} 
can be easily extended to the case of biaxial orientational 
structures. In Sec.~\ref{subsec:azim-angle},
we introduce and study 
the simplified two-dimensional (2D) diffusion model
that can be regarded as the first approximation
representing the regime of purely in-plane reorientation.

The two state and  2D diffusion models are employed 
to interpret the experimental data in Sec.~\ref{sec:results}.
Finally, in Sec.~\ref{sec:disc-concl} we present our results and
make some concluding remarks.
Technical details on solving the light
transmission problem for a biaxially anisotropic absorbing layer
and on using the analytical result to extract the extinction coefficients
from the measured dependence of absorbance on 
the incidence angle are relegated to Appendix.

\begin{figure*}[!tbh]
\centering
   \resizebox{130mm}{!}{\includegraphics*{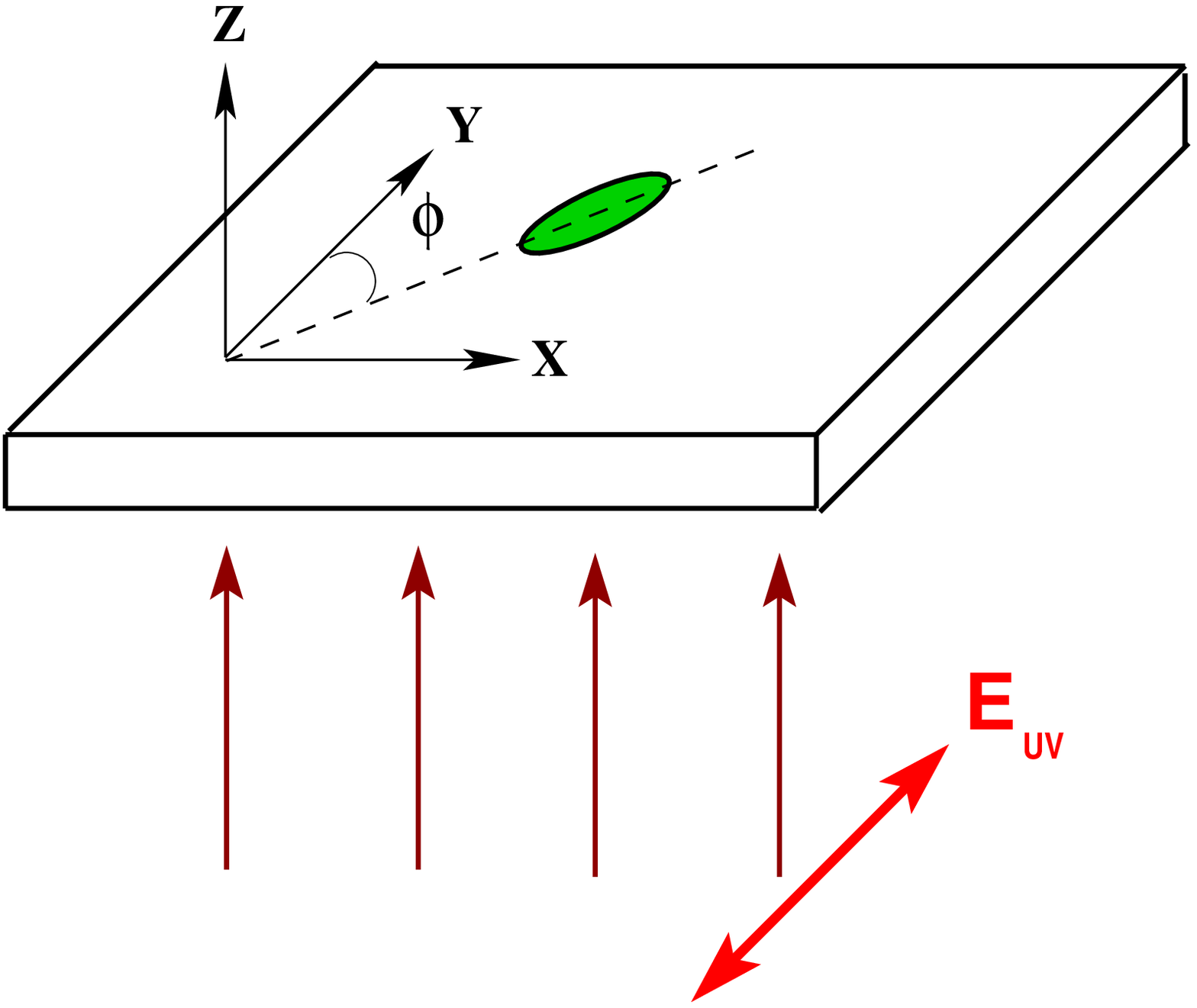}}
\caption{%
Frame of reference:
the $z$ axis is
normal to the substrate and the polarization vector of the
activating light is directed along the $y$ axis.  
}
\label{fig:frame}
\end{figure*}

\section{Master equation and two-state models}
\label{sec:m-eq-two-state}

\subsection{Order parameters, absorption tensor and biaxiality}
\label{subsec:order-param}

We assume that azo-dye molecules are cylindrically symmetric
and  orientation of a molecule in the azo-dye film
can be specified by the unit vector, 
$\uvc{u}=(\sin\theta\cos\phi,\sin\theta\sin\phi,\cos\theta)$,
directed along the long molecular axis.
Quadrupolar orientational ordering of the molecules 
is then characterized using
the  traceless symmetric second-rank tensor~\cite{Gennes:bk:1993}
\begin{equation}
  \label{eq:Q-def}
  \vc{Q}(\uvc{u})=(3\,\uvc{u}\otimes\uvc{u}-\vc{I})/2,
\end{equation}
where $\vc{I}$ is the identity matrix.

The dyadic~\eqref{eq:Q-def} averaged 
over orientation of molecules with 
the one-particle distribution function 
$\rho(\vc{r},\uvc{u})$,
describing the orientation-density
profile of azo-dye molecules,
is proportional to the \emph{order parameter tensor} $\vc{S}(\vc{r})$ 
\begin{align}
  \label{eq:avr-rho}
  \int\rho(\vc{r},\uvc{u})\vc{Q}(\uvc{u})\dd\uvc{u}=
\rho(\vc{r})\vc{S}(\vc{r}),
\end{align}
where $\dd\uvc{u}\equiv\sin\theta\dd\theta\dd\phi$,
$\rho(\vc{r},\uvc{u})=\rho(\vc{r}) f(\vc{r},\uvc{u})$,
$\rho(\vc{r})=\int\rho(\vc{r},\uvc{u})\dd\uvc{u}$
is the density profile and
$f(\vc{r},\uvc{u})$ is the normalized angular distribution. 

Throughout the paper we restrict ourselves to the case of 
spatially homogeneous systems with
$\rho(\vc{r},\uvc{u})=\rho f(\uvc{u})$.
For such systems,  
the order parameter tensor is given by
\begin{align}
&
  \label{eq:avr-Q}
  \avr{\vc{Q}}=\int \vc{Q}(\uvc{u}) f (\uvc{u})\,\dd \uvc{u}=
\vc{S},
\\
&
\vc{S}=
S (3\,\uvc{d}\otimes\uvc{d}-\vc{I})/2+
P (\uvc{m}\otimes\uvc{m}-\uvc{l}\otimes\uvc{l})/2,
  \label{eq:S}
\end{align}
where
$\lambda_1=S$, $\lambda_2=-(P+S)/2$ and 
$\lambda_3=-(\lambda_1+\lambda_2)=(P-S)/2$
are the eigenvalues of the order parameter 
tensor $\vc{S}$;
the eigenvector $\uvc{d}$ corresponding to
the largest in magnitude eigenvalue, $\lambda_1=S$, 
$|S|=\max\{|\lambda_1|,|\lambda_2|,|\lambda_3|\}$,
is the unit vector known as the \textit{director};
$P$ is the biaxiality parameter
and  the eigenvectors $\{\uvc{d},\uvc{m},\uvc{l}\}$
form a right-handed orthonormal tripod. 

In our case, the $y$ axis
is directed along
the polarization vector of the activating UV
light, $\vc{E}= E\,\uvc{y}$,
the $z$ axis is normal to the substrates
and the unit vector $\uvc{x}=[\uvc{y}\times\uvc{z}]$
is parallel to the $x$ axis (see Fig.~\ref{fig:frame}).
On symmetry grounds, it can be expected that
the basis vectors $\{\uvc{x},\uvc{y}, \uvc{z}\}$
define the principal axes of the order parameter tensor. 
So, the tensor is given by
\begin{align}
  \label{eq:S-diag}
\vc{S}= \diag(S_x,S_y,S_z)= 
S_x\, \uvc{x}\otimes\uvc{x}+
  S_y\, \uvc{y}\otimes\uvc{y}+
  S_z\, \uvc{z}\otimes\uvc{z}.
\end{align}
Then the dielectric tensor, $\bs{\varepsilon}$,
can also be written in 
the diagonal form 
\begin{align}
  \label{eq:diel-diag}
  &
\bs{\varepsilon}=\diag(\epsilon_{x},\epsilon_{y},\epsilon_{z}),
\quad
\bs{\varepsilon}_{\alpha\beta}=\epsilon_{\alpha}\,\delta_{\alpha\,\beta}.
\end{align}
In the presence of absorption,
the tensor~\eqref{eq:diel-diag} is complex-valued
and its principal values,
$\{\epsilon_{x},\epsilon_{y},\epsilon_{z}\}$,
are expressed in terms of the refractive indices,
$\{n_{x}^{(r)},n_{y}^{(r)},n_{z}^{(r)}\}$
and the extinction coefficients,
$\{\kappa_{x},\kappa_{y},\kappa_{z}\}$,
as follows~\cite{Born:bk:1999}:
\begin{align}
  \label{eq:epsil_alp}
&
\epsilon_{\alpha}=\epsilon_{\alpha}^{\prime}+ i
\epsilon_{\alpha}^{\prime\prime},
\quad
\mu\epsilon_{\alpha}=n_{\alpha}^{\,2}=(n_{\alpha}^{(r)}+
i\kappa_{\alpha})^2.  
\end{align}

We can now define
the \textit{absorption order parameters}
through the relation
\begin{align}
  \label{eq:abs_order_param}
  S_{i}^{(a)}=
\frac{2 \kappa_i - \kappa_j - \kappa_k}{2(\kappa_i+\kappa_j+\kappa_k)}=
\frac{2 D_i^{(a)} - D_j^{(a)} - D_k^{(a)}}{2(D_i^{(a)}+D_j^{(a)}+D_k^{(a)})},
\quad
i\ne j\ne k.
\end{align}
where the optical densities
$\{D_{x}^{(a)},D_{y}^{(a)},D_{z}^{(a)}\}$ are proportional
to the extinction coefficients:
$
D_{i}^{(a)}\propto \kappa_{i}. 
$
Note that the optical density
$D_{\parallel}^{(a)}\equiv D_{y}^{(a)}$  [$D_{\perp}^{(a)}\equiv D_{x}^{(a)}$] 
can be determined experimentally by measuring 
the absorption coefficient
for a testing beam 
which is
propagating along 
the normal to the film substrate
(the $z$ axis)
and is linearly polarized parallel
[perpendicular] to 
the polarization vector of the activating UV
light (the $y$ axis).

Now, following Ref.~\cite{Kis:pre:2003,Kis:pre2:2005}, 
we dwell briefly on the relation between the
orientational and the absorption order parameters 
defined in Eq.~\eqref{eq:S-diag} and Eq.~\eqref{eq:abs_order_param},
respectively.
To this end, we begin with
the absorption tensor of an azo-dye molecule 
\begin{align}
  \label{eq:sigma_mol}
 \sigma_{ij}(\uvc{u})=\sigma_{\perp}\delta_{ij}+
(\sigma_{\parallel}-\sigma_{\perp}) u_i u_j,
\end{align}
which is assumed to be uniaxially anisotropic.
Its orientational average takes the following matrix form
\begin{align}
&
  \label{eq:sigma_avr}
\avr{\bs{\sigma}}=\bigl(\sigma_{\av}\, \vc{I}+2\,
  \Delta\sigma\,\vc{S}\bigr)/3,
\\
&
\label{eq:sigma_param}
\sigma_{\av}=\sigma_{\parallel}+2\sigma_{\perp},  
\quad
\Delta\sigma=\sigma_{\parallel}-\sigma_{\perp},
\end{align}
where the angular brackets $\avr{\dots}$ denote 
orientational averaging (see Eq.~\eqref{eq:avr-Q}).

In the low concentration approximation, the optical densities 
are proportional to the corresponding 
components of the tensor~\eqref{eq:sigma_avr}
\begin{subequations}
  \label{eq:opt_dens}
\begin{align}
  \label{eq:D_par}
  D_{\parallel}^{(a)}=
&
D_{x}^{(a)}\propto \rho
\bigl(\sigma_{\av}+2\,
  \Delta\sigma\,S_{x}\bigr)/3,
\\
  \label{eq:D_perp}
  D_{\perp}^{(a)}= 
&
D_{y}^{(a)} \propto \rho
\bigl(\sigma_{\av}+2\,
  \Delta\sigma\,S_{y}\bigr)/3,
\\
&
  \label{eq:D_z}
  D_{z}^{(a)} \propto \rho
\bigl(\sigma_{\av}+2\,
  \Delta\sigma\,S_{z}\bigr)/3
\end{align}
\end{subequations}
and on substituting the expressions for the optical
densities~\eqref{eq:opt_dens} into 
Eq.~\eqref{eq:abs_order_param}
we obtain
\begin{align}
  \label{eq:abs_prop_order}
  S_{i}^{(a)}=
r_{a}\,S_i,
\end{align}
where
$r_{a}=\Delta\sigma/\sigma_{\av}=\sigma_{a}/(3+\sigma_{a})$.
So, the absorption order parameters~\eqref{eq:abs_order_param}
are equal to the corresponding elements of the order parameter 
tensor~\eqref{eq:S-diag} only
 in the limiting case 
where absorption  of waves propagating along
the long molecular axis is negligibly small: $\sigma_{\perp}\to 0$
and $\sigma_{\av}= 3 \sigma_{\perp}+\Delta\sigma\to\Delta\sigma$.
Note that the average optical density $D_x^{(a)}+D_y^{(a)}+D_z^{(a)}$
is proportional to $\rho\sigma_{\av}$ and thus typically
does not depend on the irradiation dose.

\subsection{Master equation}
\label{subsec:master-equation}

We shall assume that 
the azo-dye molecules can be 
represented by the two-level molecular systems
with the two states:
the ground state and the excited state.  
Angular distribution of the molecules in the ground state
at time $t$ is characterized by the number distribution function 
$N_{\grnd}(\uvc{u},t) = V \rho_{\grnd}(\uvc{u},t)$,
where $V$ is the volume and $\rho_{\grnd}(\uvc{u},t)$
is the corresponding one-particle distribution function. 

Similarly, the azo-dye molecules in the excited state 
are characterized by the function: $N_{\exct}(\uvc{u},t)=V \rho_{\exct}(\uvc{u},t)$.
Then the number of molecules in the ground and excited states
is given by
\begin{align}
&
  \label{eq:N_g}
  N_{\grnd}(t)\equiv N n_{\grnd}(t)=
\int N_{\grnd}(\uvc{u},t)\,\dd\uvc{u}=
V \int \rho_{\grnd}(\uvc{u},t)\,\dd\uvc{u},
\\
&
  N_{\exct}(t)\equiv N n_{\exct}(t)=
\int N_{\exct}(\uvc{u},t)\,\dd\uvc{u}=
V \int \rho_{\exct}(\uvc{u},t)\,\dd\uvc{u},
  \qquad
n_{\grnd}(t)+n_{\exct}(t)=1,
  \label{eq:consv-num}
\end{align}
where 
$N$ is the total number of molecules;
$n_{\grnd}$ and $n_{\exct}$ are the concentrations
of non-excited (ground state) and excited molecules, respectively;
$\displaystyle
\int\,\dd\uvc{u}\equiv 
\int_{0}^{2\pi}\dd\phi\int_{0}^{\pi}\,\sin\theta\,\dd\theta\,$. 

The normalized angular
distribution functions, $f_\alpha(\uvc{n},t)$, of
the ground state ($\alpha=\grnd$) and 
the excited ($\alpha=\exct$) molecules 
can be conveniently defined through the relation
\begin{equation}
  \label{eq:dis}
  N_\alpha(\uvc{u},t)\equiv V \rho_{\alpha}(\uvc{u},t) = N n_\alpha(t) f_\alpha(\uvc{u},t)\, . 
\end{equation}
linking the one-particle distribution function,
$\rho_{\alpha}$, and the corresponding concentration, $n_{\alpha}$.

The presence of long-living angular
correlations coming from anisotropic interactions between 
azo-dye molecules and collective modes of confining 
environment can be taken into account  
by using the phenomenological approach 
suggested in Refs.~\cite{Kis:eprt1:2002,Kis:pre:2003}.
In this approach,
the effective anisotropic field, that results in
the long-term stability effect 
and determines angular
distribution of the molecules in the stationary regime,
is introduced through
the additional angular distribution function,
$f_m(\uvc{u},t)$.
characterizing the additional subsystem that,
for brevity, will be referred to
as the \textit{matrix system}.   
 
It bears close
resemblance to the equilibrium distribution of the mean field theories
of photoinduced optical anisotropy~\cite{Ped:1997,Ped:1998,Hvil:2001}. 
In these theories, this distribution
has been assumed to be proportional to $\exp(-V(\uvc{u})/k_B T)$,
where $V(\uvc{u})$ is the mean-field potential that depends on the
order parameter tensor.

We shall 
write the kinetic rate equations for $N_{\alpha}(\uvc{u},t)$ in 
the general form of master equation~\cite{Gard,Kamp,Breuer:bk:2002}:
\begin{align} 
&   
\pdr{N_\alpha}{t}=
\left[
\drf{N_\alpha}{t}
\right]_{\mathrm{Diff}}+
\sum_{\beta\ne\alpha}\int
\Bigl[\,W(\alpha,\uvc{u}\,|\,\beta,\uvc{u}')
N_{\beta}(\uvc{u}',t)-
W(\beta,\uvc{u}'\,|\,\alpha,\uvc{u})\,
N_{\alpha}(\uvc{u},t)\,
\Bigr]\,\dd\uvc{u}'
\nonumber\\
&
+\gamma_{\alpha}
\Bigl[
N\,n_{\alpha}(t)
\int\Gamma_{\alpha-m}(\uvc{u},\uvc{u}') f_m(\uvc{u}',t)\dd\uvc{u}'
- N_{\alpha}(\uvc{u},t)\,\Bigr]\,,
\label{eq:master}
\end{align}
where $\alpha\,, \beta\in\{\grnd,\,\exct\}$.

The first term on the right hand side of 
Eq.~\eqref{eq:master} is due to
rotational diffusion of azo-dye molecules. 
In frictionless models this term is absent. 
It will be considered later on in Sec.~\ref{sec:f-planck-eq}.

Now we need to specify the rate of 
the $\grnd\to\exct$ transition stimulated by the incident UV light. 
For the electromagnetic wave linearly
polarized along the $y$--axis the transition rate
can be written as follows~\cite{Dum:1992,Dum:1996}:
\begin{align}
&
  \label{eq:W_g-e}
W(\exct,\uvc{u}\,|\,\grnd,\uvc{u}')=
\Gamma_{e-g}(\uvc{u},\uvc{u}')\,P_{g}(\uvc{u}'),
 \\
&
P_{g}(\uvc{u})=(\hbar\omega_t)^{-1}\Phi_{g\to e}\sum_{i,j}
\sigma_{ij}^{(g)}(\uvc{u})E_i E_j^{*}
=q_g I_{UV} (1+\sigma_a\, u_y^2)
\nonumber
\\
&
=q_g I_{UV} 
\bigl(
3+\sigma_a+ 2 \sigma_a Q_{yy}(\uvc{u})
\bigr)/3
\label{eq:P_g}
\end{align}
where 
$\bs{\sigma}^{(g)}(\uvc{u})$ is the tensor of 
absorption cross section
for the molecule in the ground state
oriented along $\uvc{u}$:
${\sigma}^{(g)}_{ij}=\sigma_{\perp}^{(g)}\delta_{ij}+
(\sigma_{||}^{(g)}-\sigma_{\perp}^{(g)})\,
u_i\,u_j$;
$
\sigma_a\equiv
(\sigma_{||}^{(g)}-\sigma_{\perp}^{(g)})/\sigma_{\perp}^{(g)}
$
is the absorption anisotropy parameter;
$\hbar\omega_t$ is the photon energy;
$\Phi_{g\to e}$ is the quantum yield of the process and 
$\Gamma_{g-e}(\uvc{u},\uvc{u}')$ describes the angular redistribution of
the molecules in the excited state;
$I$ is the pumping intensity and
$q_g\equiv (\hbar\omega_t)^{-1}\Phi_{g\to e}\sigma_{\perp}^{(g)}$.
 
Similar line of reasoning applies
to the $\exct\to\grnd$ transition to yield the expression for the rate:
\begin{align}
  \label{eq:W_e-g}
W(\grnd,\uvc{u}\,|\,\exct,\uvc{u}')=
\gamma_e\Gamma_{g-e}^{(sp)}(\uvc{u},\uvc{u}')
+q_e I_{UV}\,\Gamma_{g-e}^{(ind)\,}(\uvc{u},\uvc{u}')\,,
\end{align}
where 
$q_e\equiv (\hbar\omega_t)^{-1}\Phi_{e\to g}\sigma^{(e)}$ and
$\gamma_e\equiv 1/\tau_e$, $\tau_e$ is the lifetime of 
the excited state and the anisotropic part of the absorption cross section is
disregarded,
$\sigma_{||}^{(e)}=\sigma_{\perp}^{(e)}\equiv\sigma^{(e)}$.

Equation~\eqref{eq:W_e-g}) implies that the process of angular redistribution
for spontaneous and stimulated transitions can be different. 
All the angular redistribution
probabilities are normalized so as to
meet the standard normalization condition for
probability densities: 
\begin{equation}
  \label{eq:norm}
  \int\Gamma_{\beta-\alpha}(\uvc{u},\uvc{u}')\,\dd\uvc{u}=1\, .
\end{equation}

Using the system~\eqref{eq:master}
and the relations~\eqref{eq:W_g-e}-\eqref{eq:W_e-g} 
it is not difficult to deduce the equation for $n_{\grnd}(t)$:
\begin{equation}
  \label{eq:n-g}
  \pdr{n_{\grnd}}{t}=
\tilde{\gamma}_e\,(1-n_{\grnd})-\langle P_{g}\rangle_{\grnd}\,
n_{\grnd}\, ,
\quad
\tilde{\gamma}_e\equiv\gamma_e+q_e I_{UV},
\end{equation}
where the angular brackets $\langle\ldots\rangle_\alpha$ 
stand for averaging over the angles with the distribution function
$f_{\alpha}$ . 
Owing to the condition~\eqref{eq:norm}, this
equation does not depend on the form of the angular redistribution
probabilities.

The last square bracketed term on the right hand side
of~\eqref{eq:master} describes 
the process that equilibrates 
the absorbing molecules and the matrix system
in the absence of irradiation.
The angular redistribution probabilities
$\Gamma_{\alpha-m}(\uvc{u},\uvc{u}')$ meet the normalization
condition, so that  thermal relaxation does not
change the total fractions $N_{\grnd}$ and $N_{\exct}$. 
If there is no angular redistribution, 
then $\Gamma_{\alpha-m}(\uvc{u},\uvc{u}')=\delta(\uvc{u}-\uvc{u}')$
and both equilibrium angular distributions 
$f_{\grnd}^{(eq)}$ and $f_{\exct}^{(eq)}$ are equal to $f_m$.

The latter is the case for the mean field models 
considered in~\cite{Ped:1997,Ped:1998,Hvil:2001}.
In these models the excited molecules (\cis\ fragments)
are assumed to be long-living with $\gamma_e=0$
and $\gamma_{\grnd}=\gamma_{\exct}$. 
We can now recover the models by setting
the angular redistribution probabilities
$\Gamma_{g-e}(\uvc{u},\uvc{u}')$ and $\Gamma_{e-g}(\uvc{u},\uvc{u}')$
equal to the equilibrium distribution, $f_m=p(\uvc{u})$,
determined by the mean-field potential $V(\uvc{u})$:
$p(\uvc{u})\propto \exp(-V/k_B T)$.  
So, the mean field approach
introduces the angular redistribution operators acting as projectors
onto the angular distribution of the matrix system.  This is the
order parameter dependent distribution that characterizes 
orientation of the azo-molecules after excitation.

An alternative and a more general approach is to
determine the distribution function $f_m(\uvc{u},t)$
from the kinetic equation that can be written 
in the following form~\cite{Kis:jpcm:2002}: 
\begin{align}
&
\label{eq:gen-p}
\pdr{f_{m}(\uvc{u},t)}{t}= -\sum_{\alpha=\{\grnd,\exct\}}
\gamma_m^{(\alpha)}\,n_{\alpha}(t)\,
\Bigl[\,f_m(\uvc{u},t)
-\int\Gamma_{m-\alpha}(\uvc{u},\uvc{u}')
f_{\alpha}(\uvc{u}',t)\,\dd\uvc{u}'\,\Bigr]\,.
\end{align}

Equations for the angular distribution functions 
$f_{\grnd}(\uvc{u},t)$ and $f_{\exct}(\uvc{u},t)$ can be derived
from~\eqref{eq:master} by using
the relations~\eqref{eq:W_g-e}--\eqref{eq:n-g}.
The result is as follows
\begin{align}
&
n_{\exct}\,\pdr{f_{\exct}}{t}= 
-n_{\grnd}\,\Bigl[\,\langle P_{g}\rangle_{\grnd} f_{\exct} 
-\int\Gamma_{e-g}(\uvc{u},\uvc{u}')P_{g}(\uvc{u}')
f_{\grnd}(\uvc{u}',t)\,\dd\uvc{u}'\,\Bigr]\nonumber\\
&
-\gamma_{\exct}\,n_{\exct}\Bigl[\,f_{\exct}
-\int\Gamma_{e-m}(\uvc{u},\uvc{u}')
f_{m}(\uvc{u}',t)\,\dd\uvc{u}'\,\Bigr],
\label{eq:gen-exct}
\end{align}

\begin{align}
&
n_{\grnd}\,\pdr{f_{\grnd}}{t}= -n_{\grnd}
\left[ P_{g}(\uvc{u})-\langle P_{g}\rangle_{\grnd}
\right] f_{\grnd}
+\gamma_e\,n_{\exct}
\int\Gamma_{g-e}^{(sp)}(\uvc{u},\uvc{u}')
f_{\exct}(\uvc{u}',t)\,\dd\uvc{u}'
\nonumber\\
&
-(\gamma_e+q_e I)\,n_{\exct}\,f_{\grnd}
+q_e I n_{\exct}
\int\Gamma_{g-e}^{(ind)}(\uvc{u},\uvc{u}')
f_{\exct}(\uvc{u}',t)\,\dd\uvc{u}'
\nonumber\\
&
-\gamma_{\grnd}\,n_{\grnd}\Bigl[\, f_{\grnd}
-\int\Gamma_{g-m}(\uvc{u},\uvc{u}')
f_{m}(\uvc{u}',t)\,\dd\uvc{u}'\,\Bigr]\, .
\label{eq:gen-grnd}
\end{align}
 
The system of 
equations~\eqref{eq:n-g} and~\eqref{eq:gen-p}--\eqref{eq:gen-grnd}  
can be used as a starting point to formulate
a number of phenomenological models of POA.
We have already shown how the mean field
theories of~\cite{Ped:1997,Ped:1998,Hvil:2001} can be reformulated
in terms of the angular redistribution probabilities.

\subsection{Two-state model}
\label{subsec:2-lvl-model}

We can now describe 
our two state model.  
To this end, we follow the line of reasoning
presented in Refs.~\cite{Kis:jpcm:2002,Kis:pre:2003}. 

In this model,
the angular redistribution probabilities $\Gamma_{e-g}$
and $\Gamma_{g-e}$ are both assumed to be isotropic: 
\begin{align}
\label{eq:gamma_g_cond}
  \Gamma_{e-g}(\uvc{u},\uvc{u}')=
\Gamma_{g-e}^{(sp)}(\uvc{u},\uvc{u}')=
\Gamma_{g-e}^{(ind)}(\uvc{u},\uvc{u}')=\frac{1}{4\pi}\equiv f_{iso}.  
\end{align}
Since we have neglected anisotropy of the excited 
molecules,
it is reasonable to suppose that the equilibrium
distribution of such molecules is also isotropic,
$f_{\exct}^{(eq)}=f_{iso}\equiv (4\pi)^{-1}$, so that
\begin{align}
  \label{eq:gamma_E_cond}
\gamma_{\exct}=\gamma_{m}^{(\exct)}=0.  
\end{align}

From the other hand, we assume that
there is no angular redistribution 
\begin{align}
  \label{eq:gamma_m_cond}
\Gamma_{\alpha-m}(\uvc{u},\uvc{u}')=
\Gamma_{m-\alpha}(\uvc{u},\uvc{u}')=
\delta(\uvc{u}-\uvc{u}'),
\end{align}
and
the equilibrium angular distribution of  
molecules in the ground state is determined
by the matrix system: $f_{\grnd}^{(eq)}=f_{m}$.

Equilibrium properties of
excited and ground state molecules are thus characterized by
two different equilibrium angular distributions:
$f_{iso}$ and $f_m$, respectively. 
It means that in our model
the anisotropic field represented by $f_m$ does not influence
the angular distribution of non-mesogenic excited molecules.

\begin{subequations}
\label{eq:2lvl-ord-system-gen}
\begin{align}
  \label{eq:order_param_eqs}
    n_{\grnd}\pdr{S_{ij}^{(\grnd)}}{t}= 
&
- \frac{2\sigma_{a}}{3}\,q_{g} I_{UV}\,
n_{\grnd}\,G_{ij;\,yy}^{(\grnd)}
-\tilde{\gamma}_{e}\,(1-n_{\grnd}) 
S_{ij}^{(\grnd)} +
\notag
\\
&
+
\gamma_{\grnd}\, n_{\grnd} (S_{ij}^{(m)}-S_{ij}^{(\grnd)})\, ,
\quad
\tilde{\gamma}_e\equiv\gamma_e+q_e I_{UV},
\\
\pdr{S_{ij}^{(m)}}{t}=
&
-\gamma_{m} n_{\grnd}(S_{ij}^{(m)}-S_{ij}^{(\grnd)}),
  \label{eq:order_param_m_eqs}
\end{align}
\end{subequations}
where $G_{ij;\,mn}^{(\grnd)}$
is the order parameter correlation function given by
\begin{align}
  \label{eq:corr_funct}
  G_{ij;\,mn}^{(\grnd)}=
\langle Q_{ij}(\uvc{u})Q_{mn}(\uvc{u}) \rangle_{\grnd}-
S_{ij}^{(\grnd)}\,S_{mn}^{(\grnd)}.
\end{align}

The key point of the approach suggested in Ref.~\cite{Kis:jpcm:2002}
is the assumption that the correlators~\eqref{eq:corr_funct}
which characterize response of azo-dye to the pumping light and
enter the kinetic equations for the order parameter 
components~\eqref{eq:order_param_eqs}, 
can be expressed in terms of 
the averaged order parameters $S_{ij}^{(\grnd)}$. 

\begin{subequations}
\label{eq:2lvl-system-ord-param}
\begin{align}
n_{\grnd}\pdr{S}{t}= 
&
\frac{2\sigma_{a}}{3}\, q_g I_{UV} 
n_{\grnd}\,(5/7+2\lambda/7\,S-\lambda^2 S^2) 
\notag
\\
&
-\tilde{\gamma}_e\, (1-n_{\grnd}) S +\gamma_{\grnd} n_{\grnd} (S_m-S),
\label{eq:ord-Sx}
\\
n_{\grnd}\pdr{\Delta S}{t}= 
&
-\frac{2\sigma_{a}}{3}\, q_g I_{UV} n_{\grnd} 
\lambda (2/7+\lambda S) \Delta S 
\notag
\\
&
-
\tilde{\gamma}_e\,(1-n_{\grnd}) \Delta S +
\gamma_{\grnd} n_{\grnd} (\Delta S_m-\Delta S),
\label{eq:ord-dS}
\\
\pdr{S_m}{t}=
&
-\gamma_{m} n_{\grnd} (S_m-S)\, ,
\label{eq:ord-Sx-m}
\\
\pdr{\Delta S_m}{t}=
&
-\gamma_{m} n_{\grnd} (\Delta S_m-\Delta S)\, ,
\label{eq:ord-dS-m}
\end{align}
\end{subequations}
where $\gamma_m\equiv\gamma_m^{(\grnd)}$, 
$S\equiv S_{xx}^{(\grnd)} $,
$\Delta S\equiv S_{yy}^{(\grnd)} - S_{zz}^{(\grnd)}$,
$S_m\equiv S_{xx}^{(m)}$ and
$\Delta S_m\equiv S_{yy}^{(m)}-S_{zz}^{(m)}$.

It was shown that
the parabolic approximation used in Ref.~\cite{Kis:epj:2001}
can be improved by rescaling the
order parameter components: $S_{ii}^{(\grnd)}\to
\lambda S_{ii}^{(\grnd)}$ with
$\lambda=(1+0.6\sqrt{30})/7$ computed from the condition that
there are no fluctuations provided the molecules are perfectly aligned 
along the coordinate unit vector $\uvc{e}_i$:
$G_{ii;\,ii}^{(\grnd)}=0$ at
$S_{ii}^{(\grnd)}=1$.
In Ref.~\cite{Kis:jpcm:2002}
this heuristic procedure has also been found to
provide a reasonably accurate approximation for the correlators 
calculated by assuming that
the angular distribution of molecules can be taken in the form
of distribution functions used in the variational mean field theories
of liquid crystals~\cite{Gennes:bk:1993,Luben:bk:1995}.

\subsubsection{Long term stability and photosteady state}
\label{subsubsec:photosteady}

Mathematically, our model
is described by
equations for the order parameters 
and the concentration given in 
Eq.~\eqref{eq:2lvl-system-ord-param}
and Eq.~\eqref{eq:n-g}, respectively.
We may now pass on to discussing 
some of its general properties. 

Our first remark concerns 
the effect of the long term stability of POA.
It means that there is
the amount of the photoinduced anisotropy
preserved intact for long time
after switching off the light. 
Clearly, this is a memory effect 
and the system does not relax back to
the off state characterized by
irradiation independent equilibrium values
of the order parameters.

In order to see how this effect is described 
in our model, we assume that the activating light
is switched off at time $t=t_{\mathrm{off}}$ and 
consider subsequent evolution of 
the order parameters at $t>t_{\mathrm{off}}$.
In the absence of irradiation, 
Eq.~\eqref{eq:2lvl-system-ord-param}
decomposes into
two decoupled identical systems of equations.
for two pairs of the order parameters:
$\{S,\,S_m\}$ and $\{\Delta S,\,\Delta S_m\}$.
So, without the loss of generality 
we may restrict ourselves to the evolution
of the $x$ components of order parameters,
$\{S,\,S_m\}\equiv \{S_{xx}^{(\grnd)},\,S_{xx}^{(m)}\}$
governed by the equations
\begin{align}
&
  \label{eq:Sx_eq_off}
  \pdr{S}{t}= 
-\tilde{\gamma}_e\, (1/n_{\grnd}-1) S +\gamma_{\grnd} (S_m-S),
\\
&
  \label{eq:Sm_eq_off}
  \pdr{S_{m}}{t}= -\gamma_{m} (S_m-S)
\end{align}
supplemented with the initial conditions
\begin{align}
  \label{eq:init_cond}
  S(t_{\mathrm{off}})=S_{\mathrm{off}},\quad
  S_{m}(t_{\mathrm{off}})=S_{\mathrm{off}}^{(m)}.
\end{align}

At $I_{UV}=0$, equation for the concentration~\eqref{eq:n-g}
is easy to solve. So, for the initial value problem
with $n_{\grnd}(t_{\mathrm{off}})= n_{\mathrm{off}}$, we have
\begin{align}
  \label{eq:ng_off}
  1-n_{\grnd}= \exp(-\tilde{\gamma}_e \Delta t)
  (1-n_{\mathrm{off}}),
\quad
\Delta t = t - t_{\mathrm{off}}.
\end{align}

From Eq.~\eqref{eq:ng_off} it is clear that,
in the limiting case of
short living excited state with $\Delta t\gg 1/\tilde{\gamma}_e$,
the concentration of excited molecules, 
$n_{\exct}=1-n_{\grnd} $, rapidly decays to zero.
In this regime,
the first term on the right hand side of Eq.~\eqref{eq:Sx_eq_off}
is negligibly small and can be disregarded.
Equations~\eqref{eq:Sx_off} and~\eqref{eq:Sm_off} can now
be easily solved to yield the formulas
\begin{align}
&
  \label{eq:Sx_off}
  S(t)=
S_{\mathrm{off}}^{(\st)}+
\gamma_{\grnd}
[ S_{\mathrm{off}}
-
S_{\mathrm{off}}^{(m)}
]\exp(-\gamma\Delta t)/\gamma,
\\
&
\label{eq:Sm_off}
S_{m}(t)=
S_{\mathrm{off}}^{(\st)}-
\gamma_{m}
[ S_{\mathrm{off}}
-
S_{\mathrm{off}}^{(m)}
]\exp(-\gamma\Delta t)/\gamma,
\end{align}  
where $\gamma= \gamma_{m}+\gamma_{\grnd}$ and
\begin{align}
&
\label{eq:S_off_st}
S_{\mathrm{off}}^{(\st)}=
(
\gamma_m S_{\mathrm{off}} + 
\gamma_{\grnd} S_{\mathrm{off}}^{(m)}
)/
\gamma.
\end{align}
Evidently, the order parameters defined in Eqs.~\eqref{eq:Sx_off}
and~\eqref{eq:Sm_off} 
evolve in time approaching the stationary value~\eqref{eq:S_off_st}.
The memory effect manifests itself in 
the dependence of the stationary order parameter,  $S_{\mathrm{off}}^{(\st)}$, 
on the (initial) conditions~\eqref{eq:init_cond}
at the instant the activating light is switched off.

The photosteady states reached in the long irradiation time limit
are represented by stationary solutions of 
the system~\eqref{eq:2lvl-ord-system-gen} and 
the concentration equation~\eqref{eq:n-g}.

The steady state concentration of excited molecules 
can be expressed in terms of 
the steady state order parameter, $S_{yy}^{(\st)}$,
through the relation
\begin{equation}
1-n_{\grnd}^{(st)}=\frac{3+\sigma_{a}(1+2 S_{yy}^{(\st)})}{3(r+1)+\sigma_{a}(1+2 S_{yy}^{(\st)})}\,,
  \label{eq:ng_st}
\end{equation}
where $r\equiv \tilde{\gamma}_e/(q_g I_{UV})$.
From Eq.~\eqref{eq:ng_st} it can be seen that, in the case where
the life time of the excited state is short and the ratio $r$ is
large, $r\gg 1$, 
the fraction of the excited molecules is negligible,
so that $n_{\grnd}^{(st)}\approx 1$.

On substituting
Eq.~\eqref{eq:P_g} into the steady state relation
\begin{align}
  \label{eq:st_Pg}
  \avr{Q_{ij}(\uvc{u})P_{g}(\uvc{u})}_{\grnd}=0
\end{align}
derived from 
Eqs.~\eqref{eq:order_param_eqs} and~\eqref{eq:n-g}
we obtain equation for the steady state order parameters
\begin{align}
  \label{eq:st_Sii}
  S_{ii}^{(\st)}=-\frac{2\sigma_{a}}{3+\sigma_{a}}\,
\avr{Q_{ii}(\uvc{u})Q_{yy}(\uvc{u})}_{\grnd}.
\end{align}
From Eq.~\eqref{eq:order_param_m_eqs}
the difference between the order parameters
of the matrix system and the ground state azo-dye molecules
dies out as the photosteady state is approached,
$S_{ij}^{(\grnd)}-S_{ij}^{(m)}\to 0$ at $t\to \infty$.
Interestingly, 
Eq.~\eqref{eq:st_Sii} shows that
the order parameters, $S_{ii}^{(\st)}$, 
characterizing 
the regime of photosaturation are independent of 
the light intensity, $I_{UV}$.

For the specific form of the correlators
used to obtain the system~\eqref{eq:2lvl-system-ord-param},
the photosteady state is uniaxial with
$S_{yy}^{(\st)}= S_{zz}^{(\st)}$
and the $x$ component of the order parameter tensor,
$ S_{\st} =S_{xx}^{(\st)}$, can be found by solving
the equation
\begin{align}
  \label{eq:st_S_2lvl}
  2 \sigma_{a}\,(1/5\,+2\lambda/7\, S_{\st}-\lambda^2 S_{\st}^{\,2})
= S_{\st}(3+\sigma_{a}(1+2 S_{\st})).
\end{align}

\section{Nonlinear Fokker-Planck equations and diffusion model}
\label{sec:f-planck-eq}

\subsection{Mean field Fokker-Planck equations}
\label{subsec:mf-planck-eq}

In this section we extend the diffusion model~\cite{Chig:pre:2004}
by using the approach
based on \textit{Fokker-Planck (F-P) equations}
of the following general form~\cite{Risken:bk:1989}:
\begin{align}
  \label{eq:F-P_gen}
\pdr{P(\vc{x},t)}{t}\equiv \pdrs{t} P =
\pdrs{i} 
\left[
-D_i^{(\mathrm{drft})} P+ \pdrs{j} D_{ij}^{(\mathrm{diff})} P
\right]
=\pdrs{i} 
\left[
-\tilde{D}_i^{(\mathrm{drft})} P+ D_{ij}^{(\mathrm{diff})}\pdrs{j} P
\right],
\end{align}
where
$\pdrs{i}\equiv \pdr{}{x_i}$
and
$P(\vc{x},t)$ is the probability density (distribution function);
$D_i^{(\mathrm{drft})}$ is the drift vector and $D_{ij}^{(\mathrm{diff})}$ is the diffusion tensor.

As opposed to the linear case, in nonlinear F-P equations,
either the drift vector, $D_i^{(\mathrm{drft})}$, or the diffusion tensor,
$D_{ij}^{(\mathrm{diff})}$, depend on the distribution function, $P$:
$D_i^{(\mathrm{drft})}=D_{i}^{(\mathrm{drft})}[P]$ and 
$D_{ij}^{(\mathrm{diff})}=D_{ij}^{(\mathrm{diff})}[P]$.
The theory and applications of such equations
were recently reviewed in the monograph~\cite{Frank:bk:2005}.
Interestingly, 
according to Refs.~\cite{Curado:pre:2003,Curado:pre:2007},
nonlinear F-P equations are derived by approximating
the master equation with nonlinear effects 
introduced through the generalized transition rates.

More specifically, we concentrate on the special case of 
the so-called
\textit{
nonlinear mean-field Fokker-Planck (F-P) equations
}
\begin{align}
  \label{eq:mf_FP_gen}
  \pdrs{t} P =
\pdrs{i} 
\left\{
P\,\pdrs{i}\vdr{F[P]}{P}
\right\}
\equiv
\bnbl\cdot 
\left\{
P\,\bnbl\,
\vdr{F[P]}{P}
\right\}
\end{align}
that are characterized by the effective free energy functional,
$F[P]$.
In addition, the case of the rotational Brownian motion
will be of our primary interest.

In order to derive
the mean-field F-P equations describing 
the \textit{rotational diffusion},
the nabla operator  $-i\bnbl$ on the left-hand side of
Eq.~\eqref{eq:mf_FP_gen} , 
which is proportional to the linear momentum operator
and represent the generators of spatial translations,
should be replaced by 
the angular momentum operator 
$\JJ$ representing the generators of rotations:
$-i\bnbl\to\bs{\JJ}$~\cite{Favro:pr:1960,Doi:1988}.  
This gives the rotational F-P equation in 
the following form 
\begin{align}
  \label{eq:mf_ang_FP_gen}
    \pdrs{t} f =
- \JJ_{i} 
D_{ij}^{(\mathrm{rot})}
\left\{
f\,\JJ_{j}\,\vdr{F[f]}{f}
\right\}
\equiv
- \boldsymbol{\JJ}
\cdot 
\mvc{D}_{\mathrm{rot}}
\left\{
f\cdot\boldsymbol{\JJ}\,\vdr{F[f]}{f}
\right\},
\end{align}
where $D_{ij}^{(\mathrm{rot})}$ is the \textit{rotational diffusion tensor}
and the components of the angular momentum 
operator, $\JJ$, expressed in terms of
the Euler angles, $\bs{\omega}\equiv (\alpha, \beta,\gamma)$,  
are given by~\cite{Bie} 
\begin{subequations}
\label{eq:J_Euler}
\begin{align}
  &
  \JJ_1\equiv\JJ_{x}= -i
\Bigl\{
-\cos\alpha\,\cot\beta\,\pdrs{\alpha}
-\sin\alpha\,\pdrs{\beta}+
\frac{\cos\alpha}{\sin\gamma}\,\pdrs{\gamma}
\Bigr\},
\label{eq:Jx_Euler}
\\
&
  \JJ_2\equiv\JJ_{y}= -i
\Bigl\{
-\sin\alpha\,\cot\beta\,\pdrs{\alpha}
+\cos\alpha\,\pdrs{\beta}+
\frac{\sin\alpha}{\sin\gamma}\,\pdrs{\gamma}
\Bigr\},
\label{eq:Jy_Euler}
\\
&
  \JJ_3\equiv\JJ_{z}= -i\pdrs{\alpha}.
\label{eq:Jz_Euler}
\end{align}
\end{subequations}

When 
the effective free energy functional
is a sum of two term that represent the contributions
coming from  the \textit{effective internal energy}, $U[f]$, 
and the Boltzmann entropy term
\begin{align}
&
  \label{eq:free_en}
  F[f]=U[f]+\avr{\ln f},
\end{align}
the variational derivative of the free energy
takes the form
\begin{align}
\vdr{F}{f}=V+\ln{f} +1,
\quad
V=\vdr{U}{f},
\label{eq:deriv_fr_en}  
\end{align}
where $V$ is the \textit{mean-field potential}.

On substituting the relation~\eqref{eq:deriv_fr_en}
into Eq.~\eqref{eq:mf_ang_FP_gen}
we obtain the mean-field F-P equation
\begin{align}
   \pdrs{t} f =
- \bs{\JJ}
\cdot 
\mvc{D}_{\mathrm{rot}}
\left\{
\bs{\JJ}\,f
+
f\cdot\bs{\JJ}\,V
\right\},
  \label{eq:mf_ang_FP}
\end{align}
describing the rotational diffusion governed by 
the mean-field potential~\eqref{eq:deriv_fr_en}.
This equation can be conveniently cast into the form
\begin{align}
  \label{eq:mf_FP_D2}
   \pdrs{t} f =
D^{2}_{\JJ} f +\frac{1}{2}
\Bigl[
D^{2}_{\JJ} (f V)+
f D^{2}_{\JJ} V-
V D^{2}_{\JJ} f
\Bigr], 
\end{align}
where
the right-hand side is rewritten
using the operator 
\begin{align}
\label{eq:D2_J}
D^{2}_{\JJ} = - \JJ_{i} D_{ij}^{(\mathrm{rot})} \JJ_{j}  
\end{align}
which is quadratic in the components of 
the angular momentum operator, $\JJ_{j}$.

In the linear case where the potential, $V$,
is independent of the angular distribution function, $f$,
the F-P equation~\eqref{eq:mf_ang_FP}
has been used to study
dielectric and Kerr effect relaxation of polar liquids
based on the rotational diffusion model
the rotational motion of molecules in the presence
of external 
fields~\cite{Dejardin:jcp:1993,Dejardin:pre:2000,Kalmykov:jcp:1991,Felderhof:pre:2002,
Kalmykov:pre:2001,Kalmykov:jcp:2007}.
Rotational diffusion of a probe molecule dissolved in a 
liquid crystal phase was investigated 
in~\cite{Zannoni:jcp:1991,Zannoni:jcp:1993,Zannoni:jcp:2000}.

When molecules and the orientational distribution function 
are cylindrically  symmetric, 
the model can be described in terms the angle between
the electric field and the molecular 
axis~\cite{Dejardin:jcp:1993,Dejardin:pre:2000},
whereas angular distributions of a more general
form require using both azimuthal and polar angles
that characterize orientation of 
the molecules~\cite{Kalmykov:jcp:1991,Felderhof:pre:2002}.
In this case, for uniaxial (rod-like, calamatic) molecules, 
the distribution function $f(\alpha,\beta,\gamma)\equiv f(\bs{\omega})$
is independent of the Euler angle $\gamma$:
$f(\bs{\omega})=f(\phi,\theta)\equiv f(\uvc{u})$,
and
the angular momentum operator can be expressed in terms
of the azimuthal and zenithal (polar) angles,
$\phi$ and $\theta$, as follows
\begin{align}
  \label{eq:L_spher}
i\,\bs{\JJ}
\xrightarrow{\pdrs{\gamma}\to 0} i\,\vc{L}=[ \vc{r}\times\bnbl ]=
\uvc{e}_{\phi}\, \pdrs{\theta} - [\sin\theta]^{-1} \uvc{e}_{\theta}\, \pdrs{\phi}
\end{align}
where
\begin{subequations}
  \label{eq:spher_vectors}
\begin{align}
  &
\uvc{e}_{\theta}\equiv\uvc{e}_{x}(\uvc{r})=
\cos\theta\,\cos\phi\,\uvc{x}+
\cos\theta\,\sin\phi\,\uvc{y}-
\sin\theta\,\uvc{z},
\label{eq:e_1}
\\
&
\uvc{e}_{\phi}\equiv\uvc{e}_{y}(\uvc{r})=
-\sin\phi\,\uvc{x}+
\cos\phi\,\uvc{y},
\label{eq:e_2}
\\
 &
\uvc{r}\equiv \uvc{e}_{z}(\uvc{r})=
\sin\theta\,\cos\phi\,\uvc{x}+
\sin\theta\,\sin\phi\,\uvc{y}+
\cos\theta\,\uvc{z}.
\label{eq:e_3}
\end{align}
\end{subequations}

When the rotational diffusion tensor is 
diagonal, $D_{ij}^{(\mathrm{rot})}=D_{i}^{(\mathrm{rot})}\,\delta_{ij}$, and 
and its elements are angular independent,
 the operator~\eqref{eq:D2_J} can be written in the simplified form:
\begin{align}
\label{eq:D2_L}
D^{2}_{\JJ} = 
- \bigl(
D_{x}^{(\mathrm{rot})} \mathcal{L}_{x}^{2}+D_{y}^{(\mathrm{rot})} \mathcal{L}_{y}^{2}
+D_{z}^{(\mathrm{rot})} \mathcal{L}_{z}^{2}
\bigr).  
\end{align}
In the isotropic case with $D_{i}^{(\mathrm{rot})}=D^{(\mathrm{rot})}$,
we have
\begin{align}
\label{eq:D2_L_iso}
D^{2}_{\JJ} =
-D^{(\mathrm{rot})} \vc{L}^2= 
D^{(\mathrm{rot})} \biggl(
[\sin\theta]^{-1}\,\pdrs{\theta} (\sin\theta\,\pdrs{\theta})
+
[\sin\theta]^{-2}\,\pdrs{\phi}^2
\biggr).  
\end{align}

A more complicated biaxial case
occurs for 
asymmetric top molecules~\cite{Kalmykov:pre:2001},
macromolecules in liquid solutions~\cite{Kalmykov:jcp:2007}
and probes in the biaxial liquid crystal phase~\cite{Zannoni:jcp:1993}.
For such low symmetry,
analytical treatment cannot be simplified and 
involves the three Euler angles, 
$\bs{\omega}\equiv (\alpha, \beta,\gamma)$.

Nonlinearity in the lowest order approximation
can be introduced through
the truncated expansion for 
the internal energy functional $U[f]$
retaining one-particle (linear) and two-particle (quadratic)
terms
\begin{align}
&
  \label{eq:U_quadr}
  U[f]=\int U_1(\bs{\omega}) f(\bs{\omega})\,\dd\bs{\omega}
+\frac{1}{2}\,
\int 
f(\bs{\omega}_1) U_2(\bs{\omega}_1,\bs{\omega}_2) f(\bs{\omega}_2)
\,\dd\bs{\omega}_1\,\dd\bs{\omega}_2,
\end{align}
where
$\displaystyle
\int\dd\bs{\omega}\ldots=\int_{0}^{2\pi}\dd\alpha
\int_{0}^{\pi}\sin\beta\, \dd\beta\int_{0}^{2\pi}\dd\gamma\dots$
and 
$U_2(\bs{\omega}_1,\bs{\omega}_2)=U_2(\bs{\omega}_2,\bs{\omega}_1)$
is the symmetrized two-particle kernel.
The the effective potential
\begin{align}
&
V(\bs{\omega})
=\vdr{U}{f(\bs{\omega})}=
U_1(\bs{\omega})
+
\int
U_2(\bs{\omega},\bs{\omega}^{\,\prime}) f(\bs{\omega}^{\,\prime})\,
\dd\bs{\omega}^{\,\prime}
  \label{eq:V_quadr}  
\end{align}
is the sum of the external field potential, $U_1(\bs{\omega})$,
and the contribution coming from the two-particle 
intermolecular interactions.

For rod-like azo-dye molecules, 
the one-particle part of the effective potential~\eqref{eq:V_quadr}
can be written as a sum of the two terms:
\begin{align}
 \label{eq:U_1}
U_1(\uvc{u})=U_{I}(\uvc{u})+U_{s}(\uvc{u}), 
\end{align}
where the light-induced contribution
\begin{align}
\label{eq:U_I}
U_{I}(\uvc{u})=u_{I}\, \cnj{\vc{E}}_{UV}\cdot \vc{Q}(\uvc{u})\cdot \vc{E}_{UV}
= u_{I}\, I_{UV}\, Q_{yy}(\uvc{u})  
\end{align}
comes from the interaction of azo-molecules with the activating UV
light and the surface-induced potential
\begin{align}
\label{eq:U_s}
U_s(\uvc{u}) = u_{s}\, \uvc{z}\cdot\vc{Q}\cdot\uvc{z}
=u_{s}\, Q_{zz}(\uvc{u})
\end{align}
takes into account conditions at the bounding
surfaces of the azo-dye layer. 

Assuming that 
the two-particle interaction is of the Maier-Saupe form 
\begin{align}
  \label{eq:U_2}
  U_2(\uvc{u}_1,\uvc{u}_2)= u_2\, 
\Tr[
\vc{Q}(\uvc{u}_1)\cdot \vc{Q}(\uvc{u}_2)
]=u_2\, Q_{ij}(\uvc{u}_1)Q_{ij}(\uvc{u}_2)
\end{align}
we  derive the expression for the effective potential
of azo-dye molecules 
\begin{align}
  \label{eq:V_u}
  V(\uvc{u})=
u_{I}\, I_{UV}\, Q_{yy}(\uvc{u})+
u_{s}\, Q_{zz}(\uvc{u})+
  u_2 S_{ij}\, Q_{ij}(\uvc{u}).
\end{align}

The equilibrium angular distribution can generally be obtained as a 
stationary solution to the F-P equation~\eqref{eq:mf_ang_FP}.
It is not difficult to see that the stationary solution
given by
\begin{align}
  \label{eq:st_dist_gen}
  f_{\,\st}(\bs{\omega})= Z_{\,\st}^{-1}\,
\exp[-V(\bs{\omega})],
\quad
Z_{\st}=\int\exp[-V(\bs{\omega})]\dd\bs{\omega}
\end{align}
is the Boltzmann distribution determined by 
the effective potential.
Note that the formula~\eqref{eq:st_dist_gen}
can be obtained from the condition
\begin{align}
  \label{eq:free_en_principle}
  \vdr{F[f_{\st}]}{f(\bs{\omega})}=\mu,
\end{align}
where the constant $\mu$ can be regarded as a chemical potential
that plays the role of the Lagrange multiplier
defined through the normalization condition
$\displaystyle\int f_{\st}(\bs{\omega})\,\dd\bs{\omega}=1$. 

When the F-P equation is linear,
the stationary distribution~\eqref{eq:st_dist_gen}
describing the equilibrium state is unique.
In contrast to the linear case, the effective potential~\eqref{eq:V_u}
depends on the elements of the averaged 
orientational order parameter tensor~\eqref{eq:avr-Q}: 
$V(\bs{\omega})=V(\uvc{u}|\mvc{S})$. 
So, the components of the order parameter
tensor in the stationary state, $S_{ij}=S_{ij}^{(\st)}$,
can be found from
the self-consistency condition
\begin{align}
  \label{eq:self_cons_gen}
  S_{ij}=
\int Q_{ij}(\uvc{u})\, f_{\,\st}(\uvc{u}|\mvc{S})\,
\dd\uvc{u}.
\end{align}
In general, there are several solutions
of Eq.~\eqref{eq:self_cons_gen} 
representing multiple local extrema
(stationary points) of the free energy
\begin{align}
  \label{eq:F_st_gen}
  F[f_{\,\st}]\equiv F_{\,\st}(\mvc{S})
=
-\frac{u_2}{2}\, S_{ij} S_{ij} - \ln Z_{\,\st}(\mvc{S}).
\end{align}
Following the line of reasoning presented in 
Ref.~\cite{Frank:bk:2005} and 
using the effective free energy~\eqref{eq:free_en} 
as the Lyapunov functional,
it is not difficult to prove
the $H$ theorem for nonlinear F-P equations
of the form~\eqref{eq:mf_ang_FP_gen}.
It follows that all transient solutions converge to stationary ones 
in the long time limit.
So, each stable stationary distribution
is characterized by  the basin of attraction giving
orientational states (angular distributions)
that evolve in time approaching 
the stationary distribution.

Free energy F-P equations (both linear and non-linear) 
are generally not exactly solvable.
So, we conclude this section with remarks on
numerical methods applicable to nonlinear
F-P equations.

The method based on distributed approximating functionals (DAF)
which couples the path-integral concept to the DAF idea
is proposed for numerically solving a general class of nonlinear time-dependent
Fokker-Planck (F-P) equations
in~\cite{Zhang:pre:1997}.
The approach is applied to solve a nonlinear self-consistent dynamic
mean-field problem for which both the cumulant expansion and the
scaling theory have been found by Drozdov and Morillo~\cite{Drozdov:pre1:1996} 
to be inadequate
to describe a long-lived transient bimodality.

In Ref.~\cite{Drozdov:pre1:1996},
a finite-difference method for solving a general class of linear and
nonlinear F-P equations based on a $K$-point Stirling interpolation
formula is suggested.
A procedure to systematically evaluate all the moments of the F-P
equation by expanding them in a power series in a given function of
$t$ is suggested in~\cite{Drozdov:prl:1996}.
The methods which are extensions of this power series expansion formalism
to a general Fokker-Planck-Schr\"{o}dinger process
are presented in~\cite{Drozdov:pre:1997}. 
They are applied to a well-known problem of the decay
of a unstable state driven by exponentially correlated Gaussian noise.

\subsection{Regime of purely in-plane reorientation: 2D model}
\label{subsec:azim-angle}

In the previous section 
our model is formulated as the free energy F-P equation~\eqref{eq:mf_ang_FP}
describing rotational diffusion of azo-dye molecules
governed by the effective mean field potential~\eqref{eq:V_u}.
Since general analysis can be rather involved,
we first carefully examine our model in
the limiting two-dimensional case of 
purely in-plane reorientation.

When the symmetry is cylindrical
and the symmetry axis is directed along the normal 
to the substrates (the $z$ axis), the diffusion coefficients 
$D_x^{(\mathrm{rot})}$ and $D_y^{(\mathrm{rot})}$ are identical:
$D_x^{(\mathrm{rot})}=D_y^{(\mathrm{rot})}=D_{\perp}^{(\mathrm{rot})}$.
In this case,
the expression for the operator~\eqref{eq:D2_L}
is given by
\begin{align}
D^{2}_{\JJ} =
D_{\perp}^{(\mathrm{rot})} \biggl(
\triangle_c
+
\frac{c^2}{1-c^2}\,\pdrs{\phi}^2
\biggr)
+
D_{z}^{(\mathrm{rot})} \pdrs{\phi}^2,
\quad
c=\cos\theta,
\label{eq:D2_L_cylind}  
\end{align}
where
\begin{align}
\triangle_c
\equiv
  [\sin\theta]^{-1}\,\pdrs{\theta} (\sin\theta\,\pdrs{\theta})
=
\pdrs{c} [(1-c^2) \pdrs{c}]= (1-c^2) \pdrs{c}^2 -2 c\,\pdrs{c}. 
  \label{eq:D2_c}
\end{align}

The simplified two-dimensional model 
\begin{align}
    \pdrs{\tau} f =
&
\pdrs{\phi}
\Bigl[
\pdrs{\phi} f
+f \pdrs{\phi} V
\Bigr]
= \pdrs{\phi}^2 f +\frac{1}{2}\,
\Bigl[\,
\pdrs{\phi}^2 (f V)
\nonumber
\\
&
+ f \pdrs{\phi}^2 V-
V \pdrs{\phi}^2 f
\Bigr],
\quad
\tau = D_z^{(\mathrm{rot})} t.
 \label{eq:mf_FP_phi}  
\end{align}
immediately follows from Eq.~\eqref{eq:mf_FP_D2}
provided that
$D_{\perp}^{(\mathrm{rot})}=0$ and $f(\uvc{u},t)=f(\phi,t)$.

More interestingly, 
the F-P equation~\eqref{eq:mf_FP_phi}
can be derived by assuming that
the out-of-plane component of the unit vector
$\uvc{u}$ describing orientation of
the azo-dye molecules is suppressed
and, as is shown in Fig.~\ref{fig:frame}, 
$\uvc{u}=(\sin\phi,\cos\phi,0)$.
It implies that the molecules are constrained
to be parallel to the substrate plane
(the $x$-$y$ plane) and the orientational distribution
function takes
the factorized form
\begin{equation}
  \label{eq:suppr_out_plane}
  f(\uvc{u},t)=f(\phi,t)\,\delta(c),
\quad
c=\cos\theta,
\end{equation}
where $\delta(c)$ is the $\delta$-function.

Derivation procedure involves two steps:
(a)~substituting the relations~\eqref{eq:suppr_out_plane}
and~\eqref{eq:D2_L_cylind} into Eq.~\eqref{eq:mf_FP_D2};
and (b)~integrating the result over the out-of-plane variable $c$.
[Recall that 
$\displaystyle
\int\dd\uvc{u}=\int_{0}^{2\pi}\dd{\phi}\int_{-1}^{1}\dd c$].
The kinetic equation~\eqref{eq:mf_FP_phi} can now be readily recovered
by using the relation
\begin{align}
  \label{eq:rel_1}
  \int_{-1}^{1}\Bigl[
\triangle_c (\delta(c) V)+
\delta(c) \triangle_c V-
V \triangle_c \delta(c)
\Bigr]\dd c = 0.
\end{align}

So, in this section we consider the F-P equation~\eqref{eq:mf_FP_phi}
representing the simplest case when the out-of-plane
reorientation has been completely suppressed.
Then, for $\uvc{u}=(\sin\phi,\cos\phi,0)$, 
the diagonal elements of 
the order parameter tensor~\eqref{eq:Q-def}
averaged over the azimuthal angle
are given by
\begin{subequations}
\label{eq:order_param_azim}
\begin{align}
 &
  S_{x}=\avr{Q_{xx}}=\frac{1}{2} \avr{3\sin^2\phi -1}=
\frac{1}{4} 
\Bigl[
-3\avr{\cos 2\phi}+1
\Bigr],
\label{eq:Sx_azim}
\\
 &
  S_{y}=\avr{Q_{yy}}=\frac{1}{2} \avr{3\cos^2\phi -1}=
\frac{1}{4} 
\Bigl[
3\avr{\cos 2\phi}+1
\Bigr],
\label{eq:Sy_azim}
\\
&
S_z=\avr{Q_{zz}}=-\frac{1}{2},
\label{eq:Sz_azim}
\end{align}
\end{subequations}
where 
$\displaystyle
\avr{\ldots}=\int_{0}^{2\pi}\ldots\dd\phi$.

We can now substitute the order parameters~\eqref{eq:order_param_azim}
into the effective potential $V$ given in 
Eq.~\eqref{eq:V_u}.
The result for
the angular dependent part of
the potential is
\begin{align}
  \label{eq:V_azim}
  V=(v_1+v_2\avr{\cos 2\phi})\cos 2\phi
\equiv v\cos 2\phi,
\end{align}
where $v_1\equiv 3 u_{I} I_{UV}/4$ and
$v_2=9 u_2/8$ are the light induced and 
intermolecular interaction parameters,
respectively.

Our next step is to obtain
the system of equation for the averaged harmonics,
$c_n(\tau)=\avr{\cos n\phi}(\tau)$,
that  are proportional to the Fourier coefficients
of the distribution function, $f(\phi,\tau)$.
To this end, we integrate the F-P equation~\eqref{eq:mf_FP_phi}
multiplied by $\cos n\phi$ over the azimuthal angle
and  apply the relations 
\begin{subequations}
 \label{eq:arb_rels}
\begin{align}
 &
  \int_{0}^{2\pi}\cos n\phi\, \pdrs{\phi}^2 f\,\dd\phi
=
\avr{\pdrs{\phi}^2 \cos n\phi}= - n^2 \avr{\cos n\phi}\equiv -n^2 c_{n},
 \label{eq:arb_rels_1}
\\
&
 \int_{0}^{2\pi}\cos n\phi\, \pdrs{\phi}^2 (f V)\,\dd\phi
=
\avr{V \pdrs{\phi}^2 \cos n\phi}= - n^2 \avr{V\cos n\phi}
= -n^2 v\,(c_{n+2}+c_{n-2})/2,
 \label{eq:arb_rels_2}
\\
&
 \int_{0}^{2\pi}\cos n\phi\, f\,\pdrs{\phi}^2 V\,\dd\phi
=
\avr{\cos n\phi\, \pdrs{\phi}^2 V}=
-2 v\,(c_{n+2}+c_{n-2}),
 \label{eq:arb_rels_3}
\\
&
 \int_{0}^{2\pi}\cos n\phi\, V\,\pdrs{\phi}^2 f\,\dd\phi
=
\avr{\pdrs{\phi}^2(V\, \cos n\phi)}=
-\frac{v}{2}
\bigl[
(n+2)^2 c_{n+2}+ (n-2)^2 c_{n-2}
\bigr].
\end{align}
\end{subequations}
The resulting system reads
\begin{align}
  \label{eq:sys_c_n}
  \pdrs{\tau} c_n= - n^2 c_n + n\,v\,(c_{n+2}-c_{|n-2|}),
\quad c_{0} = 1,
\end{align}
where $v=v_1+v_2\, c_2$.

When $f(\phi+\pi)=f(\phi)$, 
the odd numbered harmonics vanish,
$c_{2 k +1}=0$.
For the even numbered harmonics,
$p_k\equiv c_{2 k}$,
the system~\eqref{eq:sys_c_n} 
can be conveniently recast into the form 
\begin{align}
& 
   \pdrs{\tau} p_k= - 4 k^2 p_k + 2 k\,v\,(p_{k+1}-p_{k-1}),
\quad
k=1,2,\ldots
\label{eq:sys_p_n}
\\
& 
p_{k}\equiv c_{2k},
\quad
p_{0} =c_{0}= 1,
\quad
v=v_1+v_2\, p_1,
\end{align}
where $p_1=\avr{\cos 2\phi}$ is 
the \textit{order parameter harmonics} 
that enter the expressions for
the orientational order parameters~\eqref{eq:order_param_azim}.

\begin{figure*}[!tbh]
\centering
   \resizebox{170mm}{!}{\includegraphics*{free_energy.eps}}
\caption{%
(a)~Intersection points represent
solutions of self-consistency
equation~\eqref{eq:self_cosist_azim}.
(b)~Stationary state free energy~\eqref{eq:F_st_gen} 
as a function of the parameter $v$.
}
\label{fig:free_ener}
\end{figure*}

\begin{figure*}[!tbh]
\centering
   \resizebox{170mm}{!}{\includegraphics*{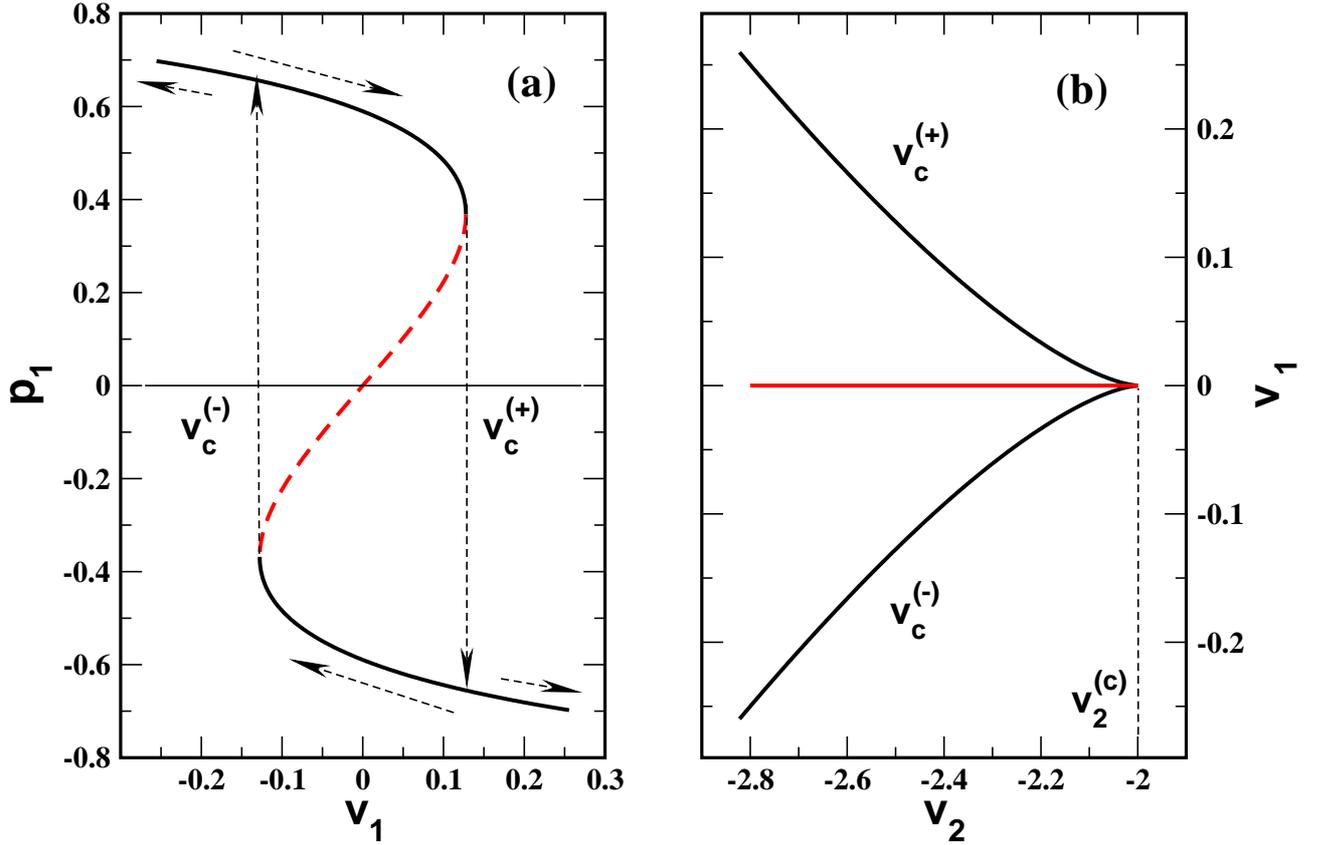}}
\caption{%
(a)~Order parameter $p_1=\avr{\cos 2\phi}$
as a function of the parameter $v_{1}$ at $v_2=-2.5$;
(b)~bifurcation curves 
in the $v_{2}$-$v_1$ plane
are typical of the cusp catastrophe 
with the cusp singularity located at $(-2,0)$.
}
\label{fig:bifur}
\end{figure*}

\begin{figure*}[!tbh]
\centering
\resizebox{140mm}{!}{\includegraphics*{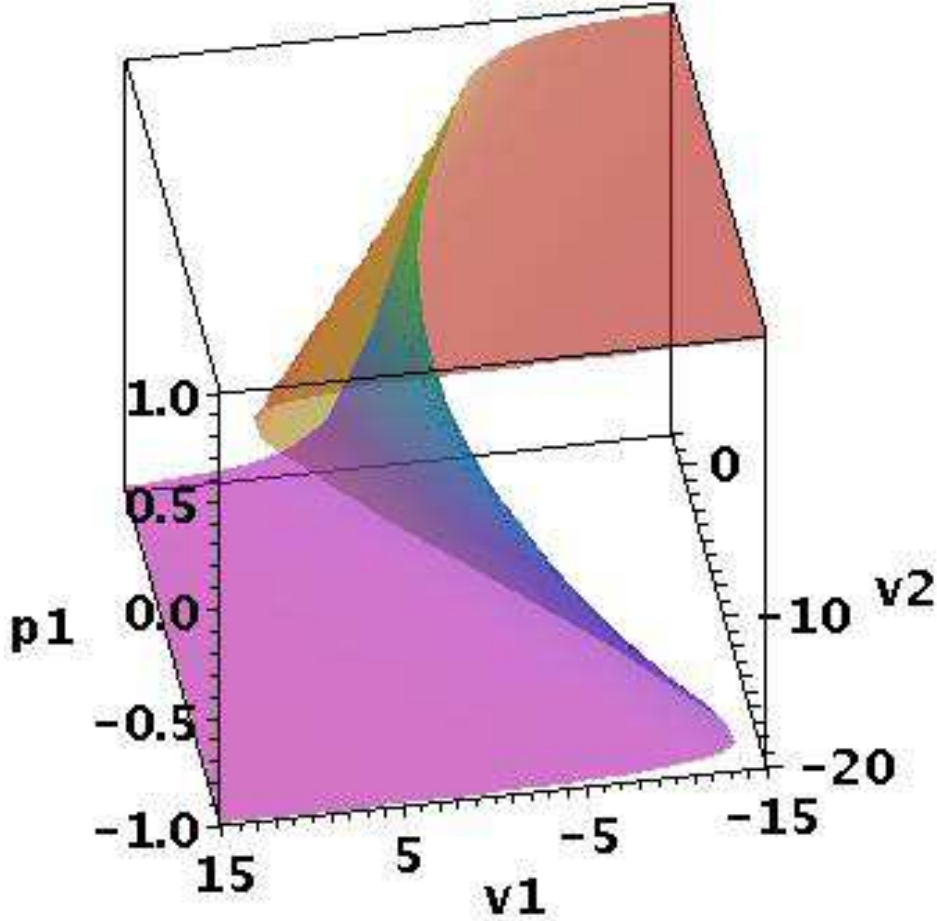}}
\caption{%
Bifurcation diagram as
the cusp surface in 
the three dimensional 
$(v_{1},v_{2},p_{1})$ space.
}
\label{fig:cusp}
\end{figure*}

\subsubsection{Bifurcations of stationary states}
\label{subsubsec:steady-state-2D}

From the general formula~\eqref{eq:st_dist_gen}
we obtain the expression
for the stationary distributions 
\begin{align}
  \label{eq:f_st_azim}
  f_{\st}=Z_{\st}^{-1}\exp[-V]=Z_{\st}^{-1}\exp[-v\cos 2\phi],
\quad
Z_{\st}=\int_{0}^{2\pi}\exp[-V]\dd\phi
\end{align}
representing the photosteady states
in the two dimensional case with the potential~\eqref{eq:V_azim}.

Equation~\eqref{eq:f_st_azim}
can now be combined with the relation
\begin{align}
  \label{eq:rel_bessel_1}
  \exp[-v\cos 2\phi]=
I_{0}(v)+
2 \sum_{k=1}^{\infty} (-1)^k I_{k}(v)\cos 2k\phi, 
\end{align}
where $I_{k}$ is 
the modified Bessel function of integer order~\cite{Abr},
to derive the formulas
\begin{align}
&
  \label{eq:Z_st_azim}
  Z_{\st}= 2\pi I_{0}(v),
\\
&
p_{k}^{(\st)}= (-1)^k I_{k}(v)/I_{0}(v).
  \label{eq:pk_st_azim}
\end{align}
giving
the stationary state statistical integral, $Z_{\,\st}$
and the averaged harmonics, $p_{k}^{(\st)}$,
expressed in terms of the parameter $v$.
Using
the recurrence relation~\cite{Abr}
\begin{align}
  \label{eq:rel_Bessel_2}
  v [ I_{k-1}(v)-I_{k+1}(v)]= 2 k I_{k}(v)
\end{align}
it is not difficult to verify 
that the formula~\eqref{eq:pk_st_azim}
gives the stationary solution to
the system~\eqref{eq:sys_p_n}
which,
in the steady state regime with $\pdrs{\tau} p_{k}=0$, 
is represented by
the finite difference equation
\begin{align}
  \label{eq:sys_pn_st}
  v \bigl(
p_{k+1}^{(\st)}-p_{k-1}^{(\st)}
\bigr)=2 k p_{k}^{(\st)}.
\end{align}

From Eq.~\eqref{eq:V_azim},
the parameter 
$v=v_{1}+v_{2} p_{1}^{(\st)}$ depends
on the order parameter harmonics
and Eq.~\eqref{eq:pk_st_azim} with $k=1$
provides the self-consistency condition 
\begin{align}
  \label{eq:self_cosist_azim}
  p_{1}^{(\st)}=(v-v_{1})/v_{2}= - I_{1}(v)/I_{0}(v).
\end{align}
This condition can also be obtained as the stationary point
equation for the stationary state free energy~\eqref{eq:F_st_gen}.
In our case, we have
\begin{align}
&
  \label{eq:U_st_azim}
  U[f_{\st}]= v p_{1}^{(\st)}-\frac{v_{2}}{2}
\bigl[
p_{1}^{(\st)}
\bigr]^2,
\\
&
  \label{eq:S_st_azim}
\avr{\ln f_{\st}}= -\ln Z_{\st}
-v p_{1}^{(\st)},
\\
&
  \label{eq:F_st_azim}
F_{\st}(v)=v_{2}^{-1}
\bigl[
-\frac{v^2}{2}+v_{1} v
\bigr] 
- \ln I_{0}(v),
\end{align}
where the additive constant is chosen so as to
have the free energy  vanishing at $v=0$. 

In Fig.~\ref{fig:free_ener}(a) it is illustrated that, 
in the $v$-$p_1$ plane,
solutions of
the self-consistency equation~\eqref{eq:self_cosist_azim}
can be found as intersection points of the curve
$- I_{1}(v)/I_{0}(v)$ and the straight line, $(v-v_{1})/v_{2}$.
It is seen that the number of the intersection points
varies between one and three
depending on the values of the parameters $v_1$ and $v_2$.

As is shown in Fig.~\ref{fig:free_ener}(b), 
for the case of three stationary states, 
the free energy curves are of the double-well potential form
with two local minima separated by the energy barrier.
From the lowest order term of the series expansion
\begin{align}
  \label{eq:p1_series_expansion}
  - I_{1}(x)/I_{0}(x)\approx
-\frac{x}{2}
+\frac{x^3}{16}
-\frac{x^5}{97}
\end{align}
it is not difficult to see that this case may occur only if
the parameter $v_2$ is less than $-2$.

Referring to Fig.~\ref{fig:free_ener}(b), 
at $v_1=0$ and $v_2<-2$, the free energy~\eqref{eq:F_st_azim}
is an even function of $v$, $F_{\st}(v)=F_{\st}(-v)$,
with two symmetrically arranged minima
representing two stable stationary states. 
When the parameter of intermolecular interaction,
$v_2$,
increases passing through its critical value,
$v_{2}^{(c)}=-2$, the minima come close together
and coalesce at the critical point. 
So, at $v_2>-2$, 
there is only one minimum
corresponding to the unique equilibrium state. 

When the activating light is switched on,
the parameter $v_1$ is distinct from zero. 
It gives rise to asymmetry effects illustrated
in Fig.~\ref{fig:free_ener}(b).
It can be seen that, at $v_1\ne 0$,
one of two minima becomes metastable. 
The local maximum representing 
the unstable stationary state
and the metastable minima merge and disappear
provided the magnitude of the parameter $v_1$
is sufficiently large.

This effect is evident from 
the curve depicted in Fig.~\ref{fig:bifur}(a)
where the stationary state 
order parameter harmonics
is plotted in the $v_1$-$p_1$ plane
by using the following parametrization
\begin{align}
  \label{eq:p1_st_v1}
p_{1}^{(\st)}=
\begin{cases}
  p_{1}=p_{1}(\xi)=- I_{1}(\xi)/I_{0}(\xi),\\
v_1=v_1(\xi)=\xi-v_2\, p_{1}(\xi).
\end{cases}
\end{align}

So, the free energy has two local minima
only if the inequalities
\begin{align}
  \label{eq:bimodal}
  v_{2}<v_{2}^{(c)}=-2,
\quad
v_{c}^{(-)}<v_{1}<v_{c}^{(+)}
\end{align}
are satisfied.
The critical values of the parameter $v_1$
depend on the reduced strength of 
intermolecular interaction $v_2$
and can be parameterized as follows
\begin{align}
  \label{eq:vc_st_v2}
v_{c}=
\begin{cases}
v_1=v_1(\xi)=\xi-v_2 p_{1}(\xi),\\
  v_{2}=v_{2}(\xi)=- [\bigl(I_{1}(\xi)/I_{0}(\xi)\bigr)'_{\xi}]^{-1}=\dfrac{1}{1+p_{1}(\xi)/\xi-p_{1}^{2}(\xi)}.
\end{cases}
\end{align}
Geometrically, in the $v_{2}$-$v_{1}$ plane,
 equation~\eqref{eq:vc_st_v2} defines
the bifurcation curves shown in Fig.~\ref{fig:bifur}(b). 
These curves form a bifurcation set 
which is the projection of the cusp surface
\begin{align}
  \label{eq:bifur_surf_param}
S_{B}=
  \begin{cases}
v_1=\xi-\zeta p_{1}(\xi),\\
v_2=\zeta,\\
  p_{1}=- I_{1}(\xi)/I_{0}(\xi)
\end{cases}
\end{align}
representing the bifurcation diagram 
in the three dimensional $(v_1,v_2,p_1)$  
space (see Fig.~\ref{fig:cusp}).
Note that the cusp bifurcation occurs as a canonical model of
a codimension 2 singularity~\cite{Kuznetsov:bk:1998}
and the surface shown in  Fig.~\ref{fig:cusp}
is typical of the cusp catastrophe~\cite{Hale:bk:1991,Hoppen:bk:2000}.

We conclude this section with the remark on how
diffusion models may account for the effect of long-term stability
by using approximation of the ``frozen'' potential 
proposed in~\cite{Chig:pre:2004}.
Mathematically, it implies that, after switching off the exciting light
at time $t=t_{\mathrm{off}}$ with
the order parameter harmonics $p_{\mathrm{off}} =p_{1}(t_{\mathrm{off}})$,
the relaxation process is governed by
the kinetic equations for the harmonics~\eqref{eq:sys_p_n}
where the parameter $v$ is changed to 
the ``frozen'' interaction parameter $v_{\mathrm{off}}=v_{2}
p_{\mathrm{off}}$.
From Eq.~\eqref{eq:pk_st_azim}
the stationary value of the order parameter harmonics
\begin{equation}
  \label{eq:p_off_st}
  p_{\mathrm{off}}^{(\st)}=-I_{1}(v_{\mathrm{off}})/I_{0}(v_{\mathrm{off}})
\end{equation}
is determined by the ``frozen'' strength of intermolecular
interaction, $v_{\mathrm{off}}$, and thus
depends on the value of the order parameter harmonics
at the time of switching, $t=t_{\mathrm{off}}$.
So,  in the two state and in 2D diffusion models
the memory effect underlying the long-term stability of POA
is described by the relations~\eqref{eq:S_off_st}
and~\eqref{eq:p_off_st}, respectively.

\section{Results}
\label{sec:results}

In Sec.~\ref{subsec:2-lvl-model}
and Sec.~\ref{subsec:azim-angle},
we employed the approaches based on 
the master and Fokker-Planck equations
to introduce two different
models:
the two state model 
and the two dimensional diffusion model,
respectively. 
In both cases, the photoinduced anisotropy is characterized
by the orientational order parameters
whose temporal evolution is governed by 
the kinetic equations of the model.

\begin{figure*}[!tbh]
\centering
   \resizebox{170mm}{!}{\includegraphics*{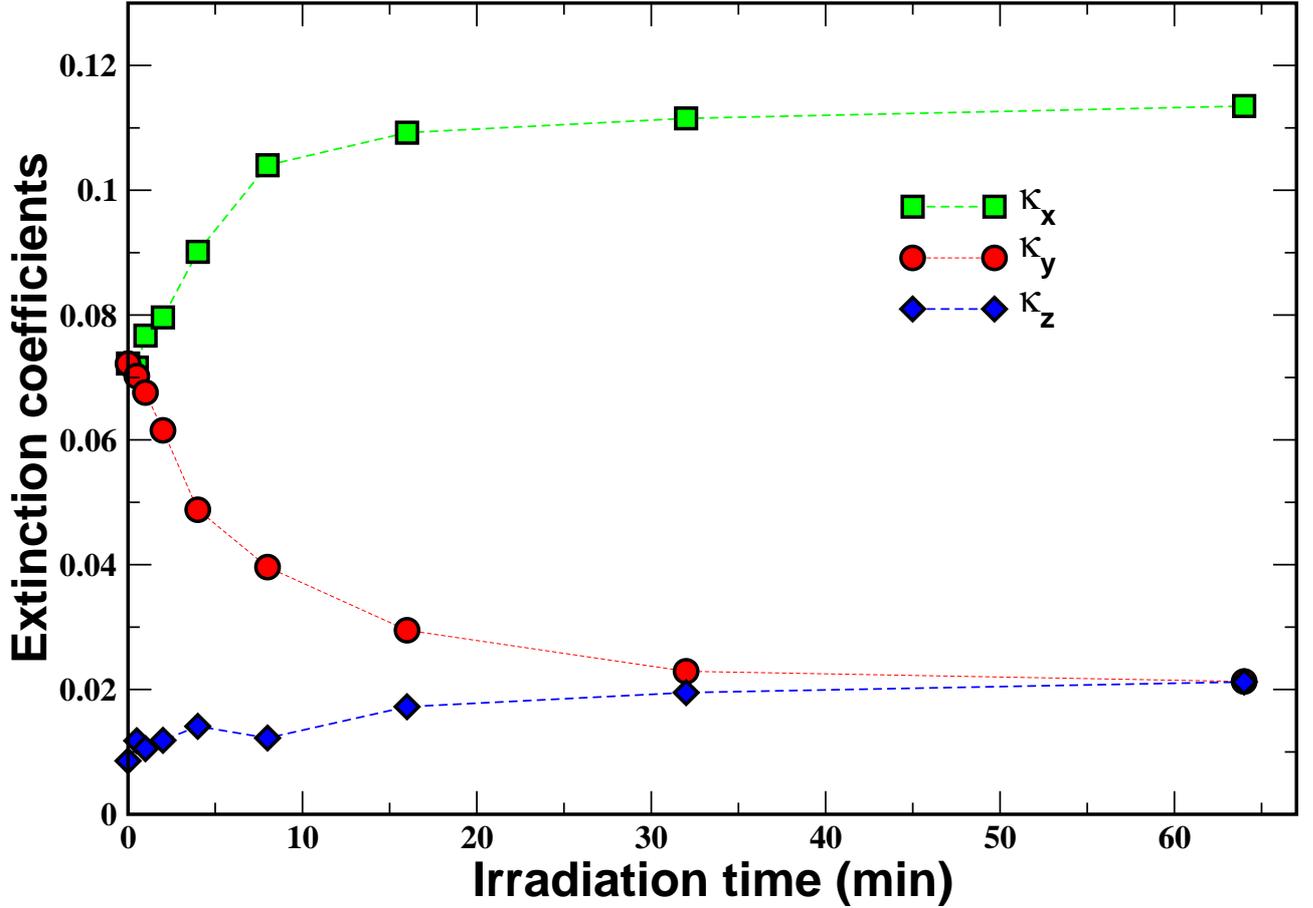}}
\caption{%
Extinction coefficients as a function
of irradiation time.
}
\label{fig:ext_coeff}
\end{figure*}

\begin{figure*}[!tbh]
\centering
   \resizebox{170mm}{!}{\includegraphics*{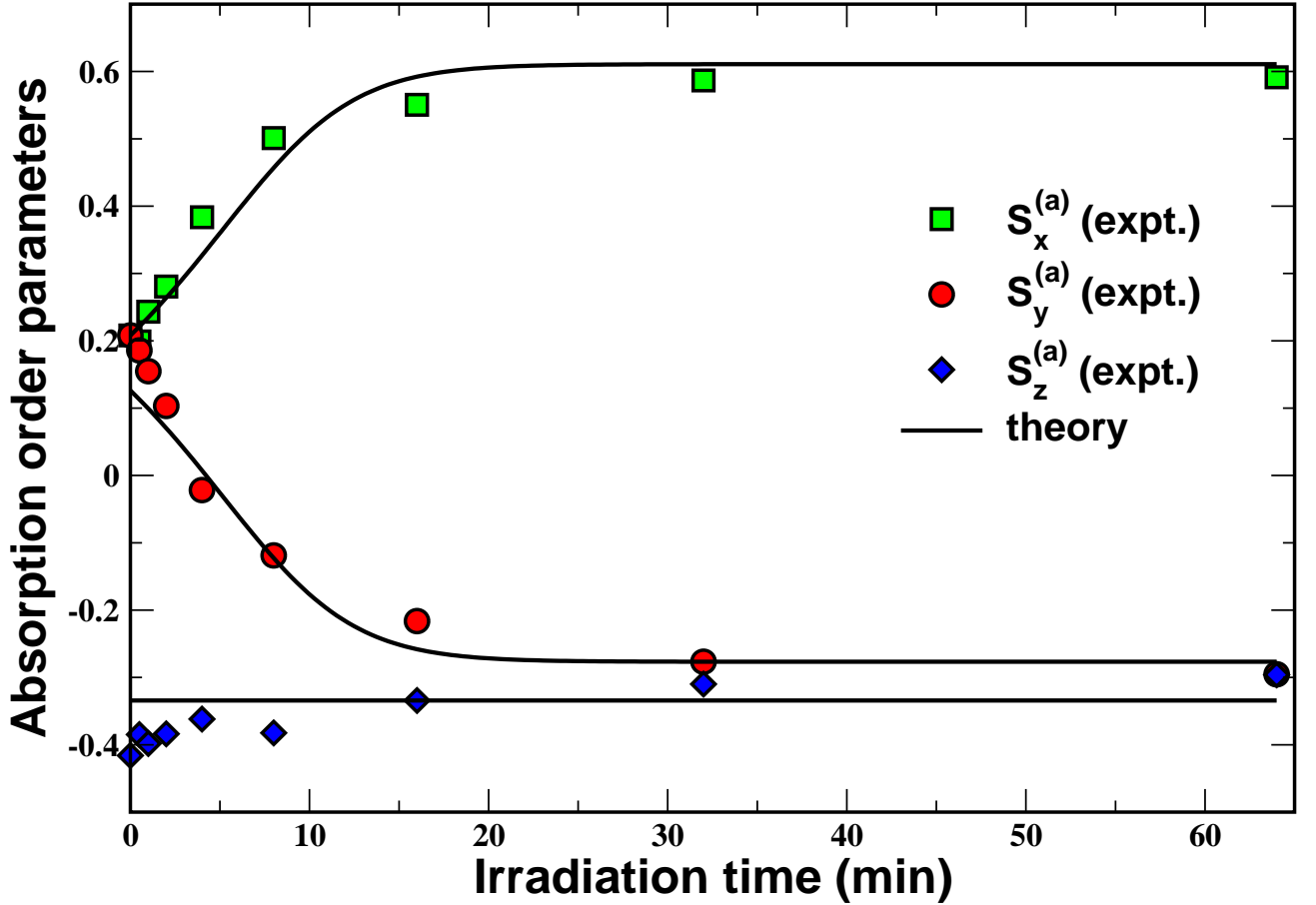}}
\caption{%
Absorption order parameters as a function
of irradiation time.
The theoretical curves are computed by
solving the system of equations~\eqref{eq:sys_p_n}
for the two dimensional diffusion model.
}
\label{fig:order_param_diff}
\end{figure*}

\begin{figure*}[!tbh]
\centering
   \resizebox{170mm}{!}{\includegraphics*{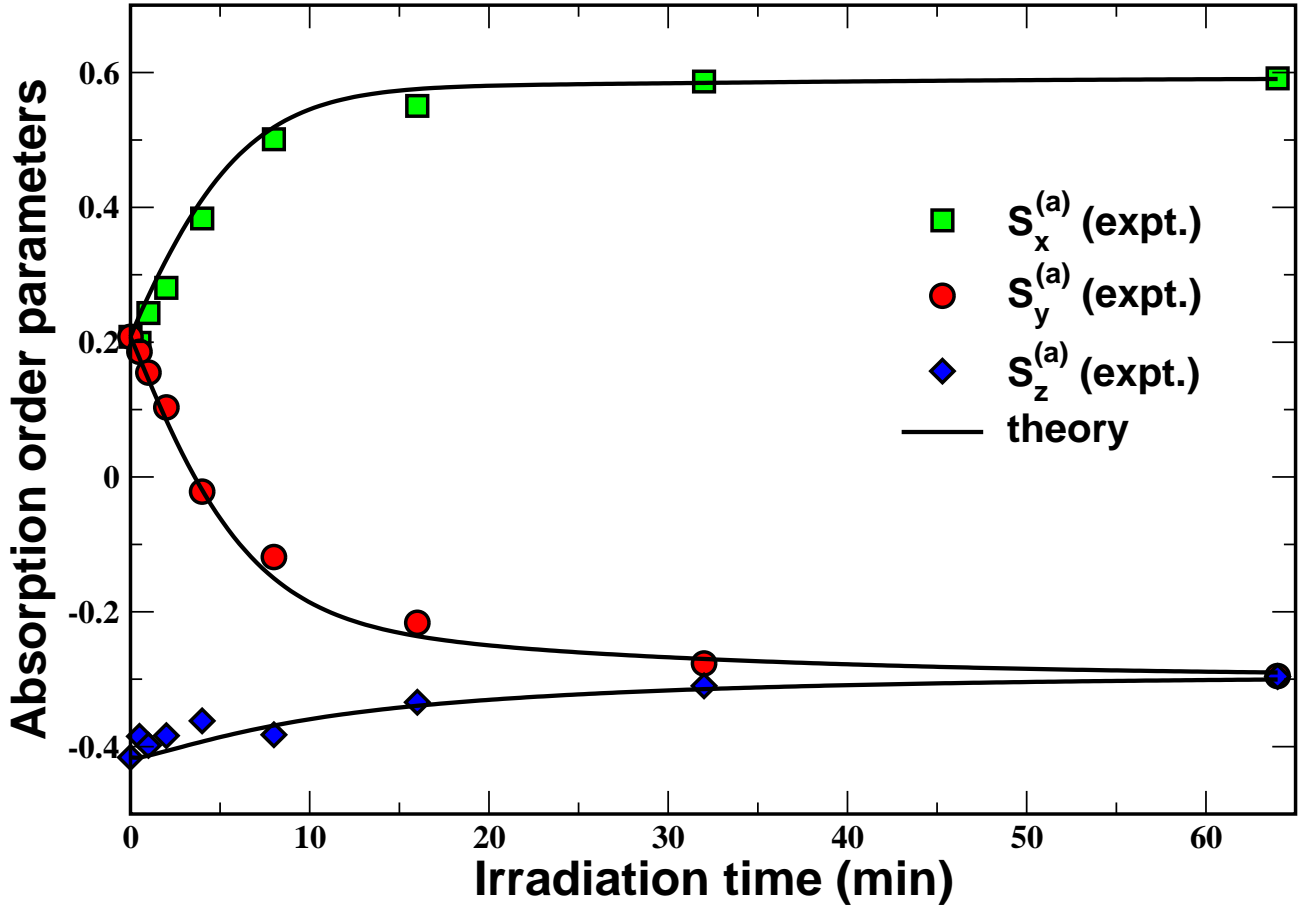}}
\caption{%
Absorption order parameters as a function
of irradiation time.
The theoretical curves are computed by
numerically solving
the kinetic equations for 
the two state model~\eqref{eq:2lvl-system-ord-param}.
}
\label{fig:order_param_2lvl}
\end{figure*}

In Sec.~\ref{subsec:order-param}, 
we discussed how
the order parameters can be related to 
absorption characteristics such as extinction 
(absorption) coefficients and optical densities.
Specifically, 
Eq.~\eqref{eq:abs_prop_order}
shows that the absorption order parameters, $S_{i}^{(a)}$,
defined in Eq.~\eqref{eq:abs_order_param}
as a function of the principal values of the extinction
coefficients, $\kappa_{i}$, are proportional
to the orientational order parameters~\eqref{eq:S-diag}.
So,  a comparison between 
the theory and experiment
can be made from
measured values of the absorption coefficients.

In thin anisotropic films,
the absorption coefficients can be determined
experimentally using the methods 
of ellipsometry~\cite{Azz:1977,Tomp:bk:2005}. 
These are generally based on the analysis
of the polarization state of light 
reflected from or transmitted through a sample. 

One of the simplest experimental procedures
is to measure the light transmittance of a film
when the testing beam is normally incident
and linearly polarized. 
Performing the measurements
for beams polarized perpendicular and parallel
to the polarization vector of the UV light
the two in-plane optical densities, $D_y^{(a)}=D_{\parallel}^{(a)}$
and $D_x^{(a)}=D_{\perp}^{(a)}$, can be obtained
as a function of the irradiation dose.

The normal component, $D_z^{(a)}$, 
then can be estimated by assuming that
the total sum of principal optical densities
\begin{align}
  \label{eq:D_tot}
  D_{tot}^{(a)}=D_x^{(a)}+D_y^{(a)}+D_z^{(a)}
\end{align}
does not depend on the irradiation dose
and the photosaturated state 
is uniaxially anisotropic with $D_y^{(a)}=D_z^{(a)}$.
More details about this approach can be found, e.g., 
in Refs~\cite{Kis:jpcm:2002,Kis:pre:2003} 
where it  was applied to azopolymer films. 

In the Appendix we show that
the absorption extinction coefficients can be 
extracted from the dependence of absorbance
on the incidence angle
measured using probe beams
which are linearly polarized parallel 
(p-polarization) and perpendicular (s-polarization)
to the plane of incidence. 
In order to fit the experimentally measured curves,
this method relies on
the analytical expressions for the transmission coefficients
of biaxially anisotropic absorbing layers
deduced in the Appendix (see Eq.~\eqref{eq:matrix-T}).

The results for the extinction coefficients,
$\kappa_{x}$, $\kappa_{y}$ and $\kappa_{z}$, 
are summarized in Figure~\ref{fig:ext_coeff} 
where the coefficients 
are plotted against the irradiation time.
The corresponding absorption order parameters,
 $S_{x}^{(a)}$, $S_{y}^{(a)}$ and $S_{z}^{(a)}$, 
evaluated from the experimental data 
by using the formula~\eqref{eq:abs_order_param}
are presented in Fig.~\ref{fig:order_param_diff} 
and Fig.~\ref{fig:order_param_2lvl}. 

It can be seen that
the initial and photosaturated states are 
both uniaxially anisotropic with $S_{x}^{(a)}=S_{y}^{(a)}$ 
and $S_{z}^{(a)}=S_{y}^{(a)}$, respectively.
So, similar to the case of azopolymers,
the transient photoinduced orientational structures 
are inevitably biaxial.

It is also clear that, before reaching the regime of photosaturation, 
the in-plane order parameters,  $S_{x}^{(a)}$ and $S_{y}^{(a)}$,
undergo pronounced changes. By contrast,
the normal component of the order parameter, $S_{z}^{(a)}$,
slowly increases
with irradiation time.
We can therefore
employ the two dimensional diffusion model 
described in Sec.~\ref{subsec:azim-angle} 
as a zero order approximation where variations of 
the normal order parameter component, $S_{z}^{(a)}$,
are neglected.

The theoretical curves shown in Figure~\ref{fig:order_param_diff}
as solid lines are computed by solving the system~\eqref{eq:sys_p_n}.
The fitting procedure is as follows.

Assuming that
the order parameter $S_{z}^{(a)}$
is constant and $S_{z}^{(a)}\approx -0.334$,
we obtain the coefficient $r_{a}$
that enter the relation~\eqref{eq:abs_prop_order},
$r_{a}\approx 0.67$, and the absorption anisotropy
parameter $\sigma_{a}\approx 6.1$. 
Then, from the experimental data, we can estimate
the order parameter harmonics in the photosteady state,
$p_{1}^{(\st)}\approx -0.887$ 
(
the corresponding in-plane order parameter is 
$S_{xx}^{(\st)}\approx 0.9$
). 
Substituting this value into the self-consistency
condition~\eqref{eq:self_cosist_azim} gives 
the equation linking the two 
dimensionless interaction parameters: $v_1$ and $v_2$.
At $v_{1}=1.0$, this equation can be solved
to yield the value of the intermolecular interaction parameter,
$v_{2}\approx -4.22$.

According to the experimental
data presented in
Figs.~\ref{fig:ext_coeff}-~\ref{fig:order_param_2lvl}, 
the irradiation time it takes to reach 
the regime of photosaturation is about $64$~min.
From the other hand,
for the computed dependence of 
the order parameter harmonics, $p_{1}$,
on $\tau\equiv D_{z}^{(\mathrm{rot})} t$, 
this regime takes place  at $\tau\ge 1.1$. 
 So, the rotational diffusion constant
$D_{z}^{(\mathrm{rot})}$
can be estimated at about $0.017$~min$^{-1}$
($\approx 2.8\times 10^{-4}$~s$^{-1}$).

Referring to Fig.~\ref{fig:order_param_diff}, 
agreement between the theoretical curves
and experiment indicates that
the two dimensional diffusion model can be regarded as a good
approximation to start from.
So,
the regime of kinetics of the photoinduced structures
in the azo-dye film appears to be
close to the limiting case of the in-plane reorientation. 

Now we consider the two state model formulated
in Sec.~\ref{subsec:2-lvl-model}.
Similar to the case of the 2D diffusion model,
our first step is  to determine 
the coefficient $r_a$ and the anisotropy parameter  
$\sigma_{a}$.
The coefficient $r_{a}$,
$r_{a}\approx 0.89$, can be calculated
as the solution of the equation 
obtained by substituting the photosaturated value
of the the absorption order parameter, 
$S_{xx}^{(a)}=r_{a} S_{\st}\approx 0.59$,  
into Eq.~\eqref{eq:st_S_2lvl}.
We also find that
the absorption anisotropy parameter
$\sigma_{a}$
is about $25.1$,
$\sigma_{a}\approx 25.1$,
and the photosteady state is characterized by
by the order parameter
$S_{\st}\equiv S_{xx}^{(\st)}\approx 0.66$.

The numerical results shown in 
Fig.~\ref{fig:order_param_2lvl} are
computed in the regime of 
photoreorientation where
the decay rate of the excited state
$\tilde{\gamma}_e\approx 2.5$~min$^{-1}$
is much larger than the excitation rate
$q_g I_{UV}/\tilde{\gamma}_e\approx 10^{-2}$,
and the thermal relaxation rates:
$\gamma_{\grnd}/\tilde{\gamma}_e\approx 5\times 10^{-3}$
and
$\gamma_{m}/\tilde{\gamma}_e\approx 2\times 10^{-2}$.
Numerical calculations in the presence of irradiation were followed by
computing the stationary values of $S$ and $\Delta S$ to which the
order parameters decay after switching off the irradiation at time
$t_{\mathrm{off}}$ 
(for more details see Sec.~\ref{subsubsec:photosteady}).

The results presented in Fig.~\ref{fig:order_param_2lvl}
suggest that the two-state model can be used to relax the assumption
on purely in-plane photoreorientation
and to go beyond the limitations of the 2D diffusion model.

\section{Discussion and conclusions}
\label{sec:disc-concl}

In order to study the kinetics of the photoinduced
ordering in azo-dye films we employed 
two different models.

The two state model is formulated by using 
the phenomenological approach developed in  
Refs.~\cite{Kis:jpcm:2002,Kis:pre:2003}.
In this approach, 
the film is represented by an ensemble of
two-level molecular systems.
So, it starts from the master equation~\eqref{eq:master}
for one-particle angular distribution functions of
the ground-state and excited molecules.
The kinetics is then determined by
the angular redistribution probabilities
that enter the photoexcitation and decay rates. 
They also define coupling between the azo-dye molecules 
and  the anisotropic field represented by the distribution function of
the matrix $f_{m}$. This anisotropic field
reflects the presence of long-living angular correlations
and stabilizes the photoinduced anisotropy. 

The resulting kinetic equations~\eqref{eq:2lvl-system-ord-param} 
for the order parameter components
are deduced by using the parabolic approximation 
suggested in Ref.~\cite{Kis:jpcm:2002} to express
the order parameter correlation functions 
in terms of the order parameter tensor.
Following the procedure described in Refs.~\cite{Kis:epj:2001,Kis:jpcm:2002},
these order parameter correlation functions
are additionally modified in order to take into account
constraints suppressing out-of-plane reorientation.
Another important assumption taken in our two state model
is that the excited molecules are isotropic and do not affect the
ordering kinetics directly. 

Similarly, there is an alternative approach 
which is formulated in Sec.~\ref{sec:f-planck-eq}
without explicit reference to excited electronic levels. 
According to this approach,
the photoinduced anisotropy arises from 
the rotational Brownian motion of azo-dye molecules
in the effective light modified potential.
Mathematically, this suggests using
the mean-field 
Fokker-Planck equation~\eqref{eq:mf_ang_FP_gen}
with the effective free energy functional~\eqref{eq:free_en}
as the equations governing the kinetics  of photoinduced ordering. 
Thus diffusion models can be defined by specifying the rotational diffusion tensor
and the effective potential~\eqref{eq:deriv_fr_en}
that enter Eq.~\eqref{eq:mf_ang_FP}.

The two dimensional model studied in Sec.~\ref{subsec:azim-angle} 
presents the simplest case to start from.
It is based on the approximation of purely in-plane photoreorientation
which assumes the normal order parameter component $S_{z}$ 
kept constant. 

In order to test applicability of this approximation,
we compared the predictions of this simple model 
with the available experimental data. 
Fro Fig.~\ref{fig:order_param_diff} it is clear that,
in azo-dye films, the kinetics of photoinduced structures
 take place in the regime close to the limiting case
of purely in-plane photoreorientation.

Referring to Fig.~\ref{fig:order_param_diff} and Fig.~\ref{fig:order_param_2lvl},
the comparison between the numerical results and the experimental data
shows that the two-state and the 2D diffusion models 
both correctly capture the basic features of 
the photo-ordering kinetics in the azo-dye layers.
It comes as no surprise that
the results computed from 
the two-level model 
give better agreement with experiment
than the ones for the 2D diffusion model.  
The primary reason for this is that the two-state model
takes into account effects due to variations of $S_{z}$.

These effects can also be taken into consideration
in the rotational diffusion approach
by expanding the orientational distribution
function into a series over 
the spherical harmonics, $Y_{lm}(\uvc{u})$ or, 
more generally, 
the Wigner $D$ functions~\cite{Bie,Devan:bk:2002}, 
$D_{m m'}^{j}(\bs{\omega})$.
The mean-field Fokker-Planck equation~\eqref{eq:mf_FP_D2}
then can be transformed into the system of nonlinear ordinary differential 
equations for the averaged harmonics, $\avr{Y_{lm}}(t)$ or $\avr{D_{m m'}^{j}}(t)$.
Equations~\eqref{eq:sys_p_n} represent
the special case of such system derived for the two dimensional model.
 
For an infinite number of equations, numerical analysis
involves truncating the system so that only a finite number
of harmonics are taken into account.
The number of harmonics is typically determined by
the required accuracy of calculations. 
Difficulties emerge if this number turns out to be very large.
For instance, this is the case for highly ordered photosteady states.

So, we have demonstrated that the phenomenological approach
of Ref.~\cite{Kis:jpcm:2002} 
and generalized diffusion models can be used as useful tools for
studying photoinduced ordering processes in azo-dye films. 
It should be noted, however, that theoretical approaches of this sort, 
by definition, do not involve explicit considerations of microscopic
details of azo-dye film physics.  
A more comprehensive study is
required to relate the effective parameters of the models and physical
parameters characterizing interactions between molecular units of films.  


\begin{acknowledgments}
This work was supported by HKUST CERG Grant No.~612406
and RPC07/08.EG01.
 \end{acknowledgments}

\appendix

\section{Transmission coefficients of biaxially anisotropic absorbing layers}
\label{sec:transm-coeff}

In this section we
derive the exact solution 
to  the transmission boundary value problem
by applying the theoretical approach developed
in Refs.~\cite{Kis:jpcm:2007,Kiselev:pra:2008} 
to the case of
biaxially (and uniformly) anisotropic absorbing layers. 

\begin{figure*}[!tbh]
\centering
\resizebox{155mm}{!}{\includegraphics*{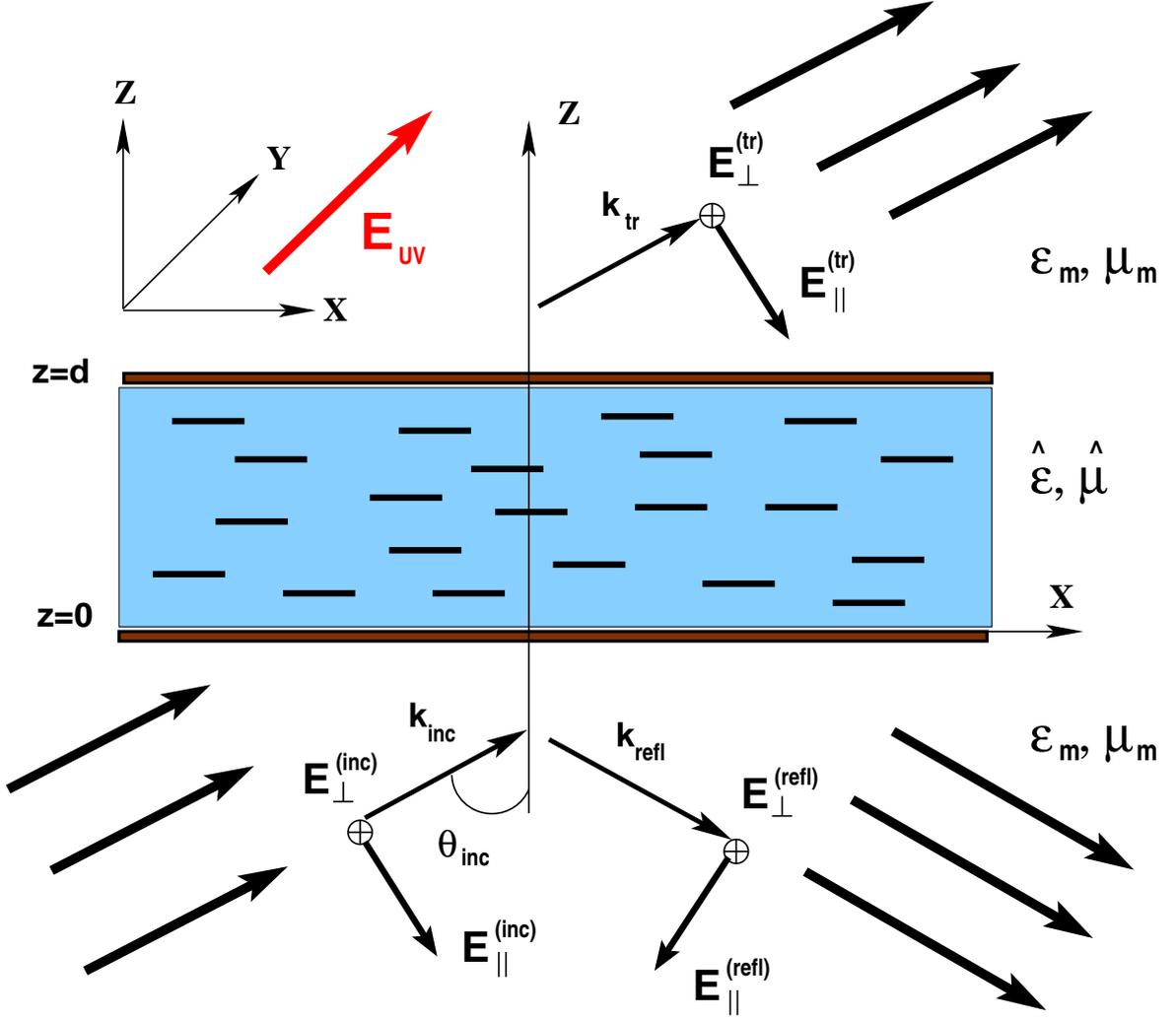}}
\caption{%
Geometry of anisotropic layer in the plane of incidence.
}
\label{fig:geom}
\end{figure*}

As is shown in Fig.~\ref{fig:geom},
we consider an absorbing uniformly anisotropic film of thickness $d$
with the $z$ axis giving the optic axis normal to the bounding
surfaces: $z=0$ and $z=d$.
The other two in-plane optic axes are assumed to be  directed along 
the unit vectors $\uvc{x}$ and $\uvc{y}$.
In this case
the dielectric tensor of the film is diagonal
and is defined in Eq.~\eqref{eq:diel-diag}.
From Eq.~\eqref{eq:epsil_alp},
the principal values of the tensor, $\epsilon_{\alpha}$,
can be expressed in terms of 
the refractive indices, $n_{\alpha}^{(r)}$,
and the extinction coefficients, $\kappa_{\alpha}$.

\begin{figure*}[!tbh]
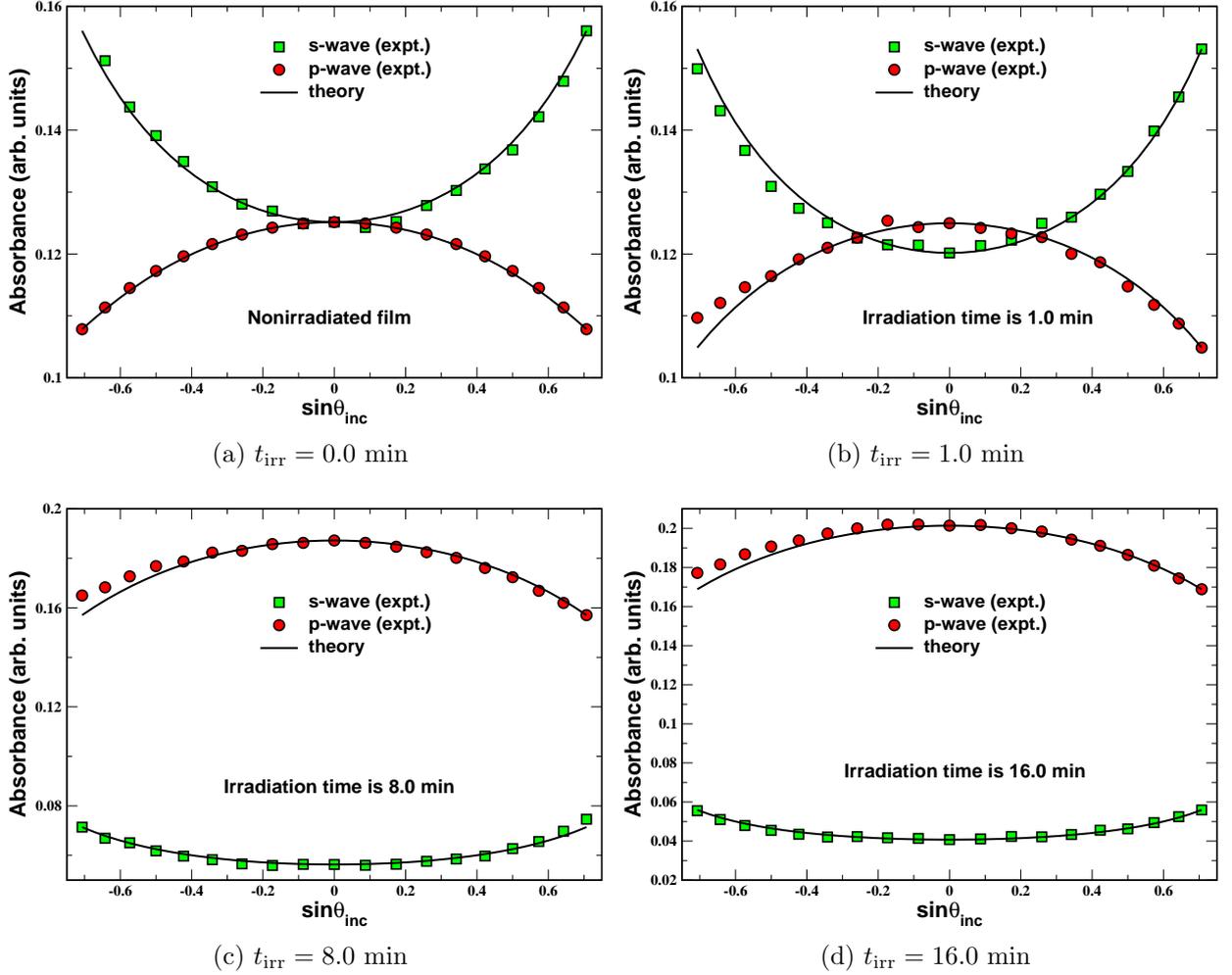

\centering
\subfloat[$t_{\mathrm{irr}}=0.0$~min ]{%
\resizebox{80mm}{!}{\includegraphics*{abs_fit_000.eps}}
\label{fig:abs_000}
}
\subfloat[$t_{\mathrm{irr}}=1.0$~min ]{%
\resizebox{80mm}{!}{\includegraphics*{abs_fit_010.eps}}
\label{fig:abs_010}
}
\\
\subfloat[$t_{\mathrm{irr}}=8.0$~min ]{%
\resizebox{80mm}{!}{\includegraphics*{abs_fit_080.eps}}
\label{fig:abs_080}
}
\subfloat[$t_{\mathrm{irr}}=16.0$~min ]{%
\resizebox{80mm}{!}{\includegraphics*{abs_fit_160.eps}}
\label{fig:abs_160}
}
\caption{%
Absorbance as a function of the incidence angle.
Circles and squares represent the data measured
for testing beam linearly polarized parallel 
(p-polarized wave) and perpendicular (s-polarized wave)
to the plane of incidence, respectively.
The theoretical curves are shown as solid lines.  
} 
\label{fig:abs_angle_fitting}
\end{figure*}

The medium surrounding the layer is assumed to be optically
isotropic and characterized by the dielectric constant $\epsilon_{\med}$
and the magnetic permittivity $\mu_{\med}$.
Referring to Fig.~\ref{fig:geom},
there are two plane waves in the half space
$z\le 0$ bounded by the entrance face of the layer:
the \textit{incoming incident wave} $\{\vc{E}_{\inc}, \vc{H}_{\inc}\}$
and the \textit{outgoing reflected wave} $\{\vc{E}_{\refl}, \vc{H}_{\refl}\}$.
In the half space $z\ge d$ after the exit face,
the only  wave is the \textit{transmitted plane wave}  $\{\vc{E}_{\trans}, \vc{H}_{\trans}\}$
which propagates along the direction of incidence
and is excited by the incident light.

So, the electric field outside the layer is
a superposition of the plane waves
\begin{subequations}
  \label{eq:E-med}
\begin{align}
&
  \label{eq:E-before}
  \vc{E}\vert_{z<0}=
\vc{E}_{\inc}(\uvc{k}_{\inc})
\ee^{i(\vc{k}_{\inc}\cdot\vc{r})}
+
\vc{E}_{\refl}(\uvc{k}_{\refl})
\ee^{i(\vc{k}_{\refl}\cdot\vc{r})},
\\
&
 \label{eq:E-after}
  \vc{E}\vert_{z>d}=
\vc{E}_{\transm}(\uvc{k}_{\transm})
\ee^{i(\vc{k}_{\transm}\cdot\vc{r})},
\end{align}
\end{subequations}
where the wave vectors
$\vc{k}_{\inc}$, $\vc{k}_{\refl}$ and $\vc{k}_{\transm}$
that are constrained to lie in the plane of incidence
due to the boundary conditions requiring
the tangential components of the electric and magnetic
fields to be continuous at the boundary surfaces.
These conditions are given by
\begin{subequations}
  \label{eq:bc-gen}
\begin{align}
&
  \label{eq:bc-E}
  \uvc{z}\times\bigl[
\vc{E}\vert_{z=0+0}-\vc{E}\vert_{z=0-0}
\bigr]
=  \uvc{z}\times\bigl[
\vc{E}\vert_{z=d+0}-\vc{E}\vert_{z=d-0}
\bigr]=0,
\\
&
 \label{eq:bc-H}
  \uvc{z}\times\bigl[
\vc{H}\vert_{z=0+0}-\vc{H}\vert_{z=0-0}
\bigr]
=  \uvc{z}\times\bigl[
\vc{H}\vert_{z=d+0}-\vc{H}\vert_{z=d-0}
\bigr]=0.
\end{align}
\end{subequations}

Another consequence of the boundary conditions~\eqref{eq:bc-gen}
is that the tangential components of the wave vectors are the same.
Assuming that the incidence plane is the $x$-$z$
plane we have
\begin{align}
  \label{eq:k-alp}
\vc{k}_{\alpha}=k_{\vac}\vc{q}_{\alpha}=k_{\med}\uvc{k}_{\alpha}=
k_x\,\uvc{x}+
k_z^{(\alpha)}\,\uvc{z},
\quad
\alpha\in\{\inc, \refl, \transm\},
\end{align}
where $k_{\med}/k_{\vac}=n_{\med}=\sqrt{\mu_{\med}\epsilon_{\med}}$
is the refractive index of the ambient medium
and
$k_{\vac}=\omega/c$ is the free-space wave number. 
The wave vector components can now be expressed
in terms of the incidence angle
$\theta_{\inc}$ as follows
\begin{align}
&
  \label{eq:k_x}
  k_x=k_{\med}\sin\theta_{\inc}\equiv k_{\vac}\, q_x,
\\
&
\label{eq:k_z-alp}
k_z^{(\inc)}=k_z^{(\transm)}=-k_z^{(\refl)}=k_{\med}\cos\theta_{\inc}\equiv
k_{\vac}\,q_{\med},
\\
&
\label{eq:qx-qm}
q_x=n_{\med}\sin\theta_{\inc},
\quad
q_{\med}=\sqrt{n_{\med}^2-q_x^2}.
\end{align}
The plane wave
traveling in the isotropic ambient medium
along the wave vector~\eqref{eq:k-alp} is transverse, so that
the polarization vector is given by
\begin{align}
&
  \label{eq:E-alp}
  \vc{E}_{\alpha}(\uvc{k}_{\alpha})=
  E_{\parallel}^{(\alpha)}\vc{e}_1(\uvc{k}_{\alpha})+
  E_{\perp}^{(\alpha)}\vc{e}_2(\uvc{k}_{\alpha}),
\\
&
\label{eq:e_x-alp}
\vc{e}_1(\uvc{k}_{\alpha})=
k_{\med}^{-1}
\bigl(
k_z^{(\alpha)}\,\uvc{x}-k_x\,\uvc{z}
\bigr),
\quad
\vc{e}_2(\uvc{k}_{\alpha})=\uvc{y},
\end{align}
where $E_{\parallel}^{(\alpha)}$ ($\equiv E_{p}^{(\alpha)}$) and
$E_{\perp}^{(\alpha)}$ ($\equiv E_{s}^{(\alpha)}$) are the in-plane and out-of-plane
components of the electric field, respectively.
The vector characterizing the magnetic field is
\begin{align}
  \label{eq:H-alp}
  \mu_{\med}\, \vc{H}_{\alpha}(\uvc{k}_{\alpha})=
\vc{q}_{\alpha}\times\vc{E}_{\alpha}(\uvc{k}_{\alpha})=
n_{\med}\,
\bigl[
  E_{\parallel}^{(\alpha)} \uvc{y}-
  E_{\perp}^{(\alpha)} \vc{e}_1(\uvc{k}_{\alpha})
\bigr],
\end{align}
where $\vc{q}_{\alpha}=k_{\vac}^{-1} \vc{k}_{\alpha}=
n_{\med}\uvc{k}_{\alpha}$.
Note that, for plane waves, the dimensionless vector
\begin{equation}
  \label{eq:q-def}
  \vc{q}=k_{\vac}^{-1}\vc{k}
\end{equation}
is parallel to $\vc{k}$ and its length gives the refractive index.
For convenience, we shall use this vector
in place of the wave vector.

The electromagnetic field of incident,
transmitted and reflected waves propagating in the ambient medium
is of the general form
\begin{align}
  \label{eq:EH-form}
  \left\{
\vc{E}, \vc{H}
\right\}
=
\left\{
\vc{E}(z), \vc{H}(z)
\right\}
\ee^{i ({k}_{x} x -\omega t)}.
\end{align}
On substituting the relations~\eqref{eq:EH-form}
into the Maxwell equations 
we can obtain the equations for 
the tangential components of the electromagnetic field
inside the anisotropic layer.
The result can be written  
in the following $4\times 4$ matrix 
form
\begin{align}
  \label{eq:matrix-system}
  -i\pdrs{\tau}\vc{F}=\mvc{M}\cdot\vc{F}\equiv
    \begin{pmatrix}\mvc{M}_{11}&\mvc{M}_{12}\\\mvc{M}_{21}&\mvc{M}_{22} \end{pmatrix}
    \begin{pmatrix}\vc{E}_{P}\\\vc{H}_{P} \end{pmatrix},
\quad
\tau\equiv k_{\vac} z,
\end{align}
where
$
\vc{E}_{P}(z)=\begin{pmatrix}
E_x(z)\\ E_y(z)
\end{pmatrix}
$
and
$
\vc{H}_{P}(z)=\begin{pmatrix}
H_y(z)\\ -H_x(z)
\end{pmatrix}
$.

For the dielectric tensor~\eqref{eq:diel-diag}
with the plane of incidence parallel to the $x$-$z$ plane,
from the general expressions derived in 
Refs.~\cite{Kis:jpcm:2007,Kiselev:pra:2008},
the $2\times 2$ matrices $\mvc{M}_{ij}$ characterizing 
the block structure of the matrix $\mvc{M}$ are given by
\begin{align}
  \label{eq:matrix-M}
  &
\mvc{M}_{12}=
\epsilon_{z}^{-1}
\begin{pmatrix}
  n_z^{\,2}-q_x^{\,2} & 0\\
0 & n_z^{\,2}
\end{pmatrix},
\quad
\mvc{M}_{11}=\mvc{0},
\\
&
\mvc{M}_{21}=
\mu^{-1}
\begin{pmatrix}
  n_x^{\,2}     & 0\\
0 & n_y^{\,2}-q_x^{\,2}
\end{pmatrix},
\quad
\mvc{M}_{22}=\mvc{0}.
\end{align}

According to the computational procedure
developed in Refs.~\cite{Kis:jpcm:2007,Kiselev:pra:2008},
the transmission and reflection matrices
defined through the linear input-output relations
\begin{align}
  \label{eq:transm-rel}
  \begin{pmatrix}
E_{\parallel}^{(\transm)}\\
E_{\perp}^{(\transm)}
\end{pmatrix}
&
=\mvc{T}
\begin{pmatrix}
E_{\parallel}^{(\inc)}\\
E_{\perp}^{(\inc)}
\end{pmatrix},
\quad
 \begin{pmatrix}
E_{\parallel}^{(\alpha)}\\
E_{\perp}^{(\alpha)}
\end{pmatrix}
\equiv
\begin{pmatrix}
E_{p}^{(\alpha)}\\
E_{s}^{(\alpha)}
\end{pmatrix},
\\
  \label{eq:refl-rel}
  \begin{pmatrix}
E_{\parallel}^{(\refl)}\\
E_{\perp}^{(\refl)}
\end{pmatrix}
&
=
\mvc{R}
\begin{pmatrix}
E_{\parallel}^{(\inc)}\\
E_{\perp}^{(\inc)}
\end{pmatrix}
\end{align}
can be expressed
in terms of the linking matrix
\begin{align}
&
  \label{eq:W-op}
  \mvc{W}=
\mvc{V}_{\med}^{-1}\cdot\mvc{U}^{-1}(h)\cdot\mvc{V}_{\med}=
\begin{pmatrix}
\mvc{W}_{11} & \mvc{W}_{12}\\
\mvc{W}_{21} & \mvc{W}_{22}
\end{pmatrix}
\end{align}
as follows
\begin{align}
 &
  \label{eq:trans-mat1}
 \mvc{T}
=\mvc{W}_{11}^{-1},
\\
&
  \label{eq:refl-mat1}
 \mvc{R}=\mvc{W}_{21}\cdot\mvc{W}_{11}^{-1}=\mvc{W}_{21}\cdot\mvc{T}.
\end{align}

The expression for the linking matrix~\eqref{eq:W-op}
involves the inverse of the evolution operator
\begin{align}
  \label{eq:evol_oper}
     \mvc{U}^{-1}(h)=\mvc{U}(-h)=\exp\{-i \mvc{M}\, h\},
\quad
h=k_{\vac}d
\end{align}
and the eigenvector matrix for the ambient medium
\begin{align}
  \label{eq:Vm-block}
  \mvc{V}_{\med}=
\begin{pmatrix}
\mvc{E}_{\med} & -\bs{\sigma}_3 \mvc{E}_{\med}\\
\mvc{H}_{\med} & \bs{\sigma}_3 \mvc{H}_{\med}\\
\end{pmatrix}
\end{align}
which is characterized by the two diagonal $2\times 2$ matrices
\begin{align}
&
\label{eq:HE_med}
  \mvc{E}_{\med}=\diag(q_{\med}/n_{\med}, 1),
\quad
  \mu_{\med}\,\mvc{H}_{\med}=\diag(n_{\med}, q_{\med}),
\end{align}
where $\bs{\sigma}_3=\diag(1,-1)$.

In our case, the resulting expression for the evolution operator
is
\begin{align}
  \label{eq:matrix-U}
     \mvc{U}(h)=\exp\{i \mvc{M}\, h\}=\mvc{V}\cdot
      \begin{pmatrix}
        \mvc{U}_{+}&\mvc{0}\\
\mvc{0}&\mvc{U}_{-}
      \end{pmatrix}
\mvc{N}^{-1}\cdot\tcnj{\mvc{V}}\cdot\mvc{G}
\,,\quad
\mvc{G}=
\begin{pmatrix}
  \mvc{0}&\mvc{I}_2\\
\mvc{I}_2&\mvc{0}
\end{pmatrix},
\end{align}
where $\mvc{I}_2=\diag(1,1)$ and
\begin{align}
&
  \label{eq:matrix-Ud}
\mvc{U}_{\pm}
=
\exp\{\pm i \mvc{Q}\, h\},
\quad
\mvc{Q}
=
\begin{pmatrix}
  q_{p} &0\\
0& q_{s} 
\end{pmatrix},
\\
&
\label{eq:q_ps}
q_p=\frac{n_x}{n_z}\sqrt{n_z^2-q_x^2},
\quad
q_s=\sqrt{n_y^2-q_x^2},
\\
&
  \label{eq:matrix-VN}
  \mvc{V}
=
\begin{pmatrix}
  \mvc{E} & -\bs{\sigma}_{3}\mvc{E}\\
  \mvc{H} & \bs{\sigma}_{3}\mvc{H}
\end{pmatrix},
\quad
\mvc{N}=\frac{2}{\mu}\diag(q_p,q_s,-q_p,-q_s),
\\
&
\label{eq:matrix-E}
\mvc{E}=
\begin{pmatrix}
  q_p/n_x & 0\\
  0 & 1
\end{pmatrix},
\quad
\mvc{H}=
\frac{1}{\mu}
\begin{pmatrix}
  n_x & 0\\
  0 & q_s
\end{pmatrix}.
\end{align}

We can now substitute the operator~\eqref{eq:matrix-U}
into the linking matrix~\eqref{eq:W-op}
and obtain the
transmission and reflection matrices 
using the relations~\eqref{eq:trans-mat1}
and~\eqref{eq:refl-mat1}.
The result is given by 
\begin{align}
&
  \label{eq:matrix-T}
  \mvc{T}=
\begin{pmatrix}
t_{p}(q_x) & 0\\
0& t_{s}(q_x)
\end{pmatrix}
=
\frac{\mvc{I}_2-\mvc{P}^2}{\mvc{I}_2-\mvc{U}_{+}^2\mvc{P}^2}\,\mvc{U}_{+},
\\
&
  \label{eq:matrix-R}
  \mvc{R}=
\begin{pmatrix}
r_{p}(q_x) & 0\\
0& r_{s}(q_x)
\end{pmatrix}
=
\bs{\sigma}_3\frac{\mvc{I}_2-\mvc{U}_{+}^2}{\mvc{I}_2-\mvc{U}_{+}^2\mvc{P}^2}\,\mvc{P},
\\
&
\label{eq:matrix-P}
\mvc{P}=\mvc{V}_{-}\,\mvc{V}_{+}^{-1},
\quad
\mvc{V}_{\pm}
=
\begin{pmatrix}
  \dfrac{n_x}{\mu n_{\med}} q_{\med}\pm
  \dfrac{n_{\med}}{\mu_{\med}n_x} q_{p}
&0\\
0&\mu^{-1} q_s\pm \mu_{\med}^{-1} q_{\med}
\end{pmatrix}.
\end{align}
From Eq.~\eqref{eq:matrix-T}
and Eq.~\eqref{eq:matrix-R},
non-diagonal elements of
both transmission and reflection matrices vanish.
Algebraically, this is a consequence of the diagonal form
of the block matrices that enter the operator of
evolution~\eqref{eq:matrix-U}.

So, absorption of plane waves linearly polarized parallel
and perpendicular to the plane of incidence
can be characterized by the effective optical densities,
$D_p$ and $D_s$, expressed in terms of the corresponding
transmission coefficients:
\begin{align}
  \label{eq:Ds_Dp}
  D_{p,\,s}(\theta_{\inc})\equiv
  D_{p,\,s}(q_x)=-2\ln|t_{p,\,s}(q_x)|.
\end{align}
The optical densities~\eqref{eq:Ds_Dp} are proportional to 
the absorbances measured experimentally,
$D_p^{(\mathrm{exp})}$ and $D_s^{(\mathrm{exp})}$,
and determine the theoretical dependence of the absorbance 
on the incidence angle, $\theta_{\inc}$
(or, equivalently, on the incidence angle parameter 
$q_x=n_{\med}\sin\theta_{\inc}$).

In Fig.~\ref{fig:abs_angle_fitting},
the experimental data 
on angular dependence of absorbance
measured in the azo-dye SD1 film of the thickness 15~nm 
at different irradiation doses
are fitted by the theoretical curves 
computed from the formula~\eqref{eq:Ds_Dp}.

For our purposes, full description of a rather standard
experimental procedure is not important 
(more details can be found 
in~\cite{Kiselev:idw:2008}).
So, without going into details we note that 
the film was illuminated with linearly polarized
UV light at varying exposure time
by using LED exposure light source.
The wavelength and the intensity of the actinic light
were 365~nm and 3.0~mW/cm$^2$, respectively.

In Fig.~\ref{fig:ext_coeff},
the extinction coefficients of the azo-dye layer
found as the fitting parameters
are plotted as a function of the irradiation time.


\end{document}